\documentclass[aps,preprint,pre,superscriptaddress]{revtex4-1}

\usepackage{graphicx}		% Einbinden von Grafiken
\usepackage{color}
\usepackage{amsmath}
\usepackage{amssymb}
\usepackage{bm}
\usepackage[normalem]{ulem} % either use this (simple) or
\usepackage{soul} % use this (many fancier options)
\usepackage{ulem}
\usepackage{textcomp}
\usepackage{siunitx}

\usepackage[hidelinks]{hyperref}

\begin{document}

\title{Integer topological defects of cell monolayers - mechanics and flows}

\author{Carles Blanch-Mercader}
\affiliation{Department of Biochemistry, University of Geneva, 1211 Geneva, Switzerland}
\affiliation{Department of Theoretical Physics, University of Geneva, 1211 Geneva, Switzerland}
 
\author{Pau Guillamat}
\affiliation{Department of Biochemistry, University of Geneva, 1211 Geneva, Switzerland}

\author{Aur\'elien Roux}
\affiliation{Department of Biochemistry, University of Geneva, 1211 Geneva, Switzerland}

\author{Karsten Kruse}
\affiliation{Department of Biochemistry, University of Geneva, 1211 Geneva, Switzerland}
\affiliation{Department of Theoretical Physics, University of Geneva, 1211 Geneva, Switzerland}
\affiliation{NCCR Chemical Biology, University of Geneva, 1211 Geneva, Switzerland}

%\pacs{87.17.Jj, 47.20.Ma, 47.20.Ky, 87.17.Rt}

%\date{\today}

\begin{abstract}
Monolayers of anisotropic cells exhibit long-ranged orientational order and topological defects. During the development of organisms,  orientational order often influences morphogenetic events. However, the linkage between the mechanics of cell monolayers and topological defects remains largely unexplored. This holds specifically at the time scales relevant for tissue morphogenesis. Here, we build on the physics of liquid crystals to determine material parameters of cell monolayers. In particular, we use a hydrodynamical description of an active polar fluid to study the steady-state mechanical patterns at integer topological defects. Our description includes three distinct sources of activity: traction forces accounting for cell-substrate interactions as well as anisotropic and isotropic active nematic stresses accounting for cell-cell interactions. We apply our approach to C2C12 cell monolayers in small circular confinements, which form isolated aster or spiral topological defects. By analyzing the velocity and orientational order fields in spirals as well as the forces and cell number density fields in asters, we determine mechanical parameters of C2C12 cell monolayers. Our work shows how topological defects can be used to fully characterize the mechanical properties of biological active matter.
\end{abstract}

\maketitle

\section{Introduction}

Collective cell migration plays a major role in the regulation of vital biological processes, including tissue morphogenesis, wound healing, and tumor progression~\cite{Ladoux2017,Friedl2009,Hakim2017}. Cell migration is driven by the cytoskeleton, a network of multiple protein filaments, such as actin, and molecular motor complexes, such as myosin. As an active material, the cytoskeleton can generate mechanical stresses at the cellular level by consuming the chemical fuel Adenosine-Triphosphate (ATP). Cell-cell junctions can transmit such mechanical stresses to neighboring cells, which leads to collective cell migration. 

During morphogenesis and regeneration, cells commonly display anisotropic distributions of intracellular constituents. Examples are stress fibers, which are bundles of actin filaments and myosin motors. In cells, these structures can organize into phases with orientational order~\cite{Dalby2002,Prager-Khoutorsky2011,Gupta2019}. Other forms of orientational cellular order are resulting from the symmetry breaking between front and back of migrating cells. At the front, migration is generated by a distinct structure enriched with branching actin filaments called the lamellipodium. 

Physical interaction between such anisotropic cells can lead to long-range orientational order with varying degrees of symmetry. For instance, polarity markers in mouse liver or confluent monolayers of fibroblasts \textit{in vitro} exhibit nematic order~\cite{Morales-Navarrete2019,Duclos2014}. Similar to liquid crystals~\cite{DeGennes1995}, nematic refers to order that is invariant under inversions of the cell orientation. Signatures of polar order, where this invariance is absent, have been reported in spreading epithelial monolayers~\cite{Farooqui2005,Trepat2009,Reffay2011,Peyret2019}. 

Orientational fields exhibit topological defects, where the orientation is not well-defined. These defects are characterized by their topological charge, which is determined by counting the number of rotations the orientational field performs when following a closed trajectory around the defect center~\cite{DeGennes1995}. Polar order fields can present topological defects with an integer charge, whereas nematic order fields can also exhibit half-integer defects. In active materials, the characteristics of the mechanical patterns around topological defects depend on details of the underlying active processes. In particular, studying the dynamics of half-integer topological defects, one can infer whether the active stresses are contractile or extensile~\cite{Sanchez2012,Saw2017,Kawaguchi2017,Duclos2018,Blanch-Mercader2018m,Copenhagen2020}.

Several theoretical studies suggest that in active systems, well-defined mechanical patterns and flows can emerge around topological defects~\cite{Giomi2013,Giomi2014,Thampi2014,Shankar2018,Hoffmann2020}. Based on this idea, one can qualitatively understand the structure of collective flows of active systems, such as purified cytoskeletal motor-filament suspensions, by considering the dynamics of topological defect assemblies~\cite{Sanchez2012,Guillamat2016,Guillamat2017,Hardouin2019,Opathalage2019a}. Similar ideas were applied to multicellular systems to interpret various processes including cell extrusion~\cite{Saw2017}, changes in cell density~\cite{Kawaguchi2017}, or morphogenetic events during the regeneration of the freshwater polyp \textit{hydra}~\cite{Livshits2017,Maroudas-Sacks2020}. These findings suggest that orientational fields can organize cell stress patterns and guide collective cell migration.

In this work, we show that the dynamics of individual topological defects can be used to determine mechanical properties of active systems. To this end, we first develop a hydrodynamic approach to study the forces, orientation, and flows around integer topological defects in compressible active fluids. Our phenomenological description accounts for three types of active processes, corresponding to polar cell-substrate forces as well as isotropic and anisotropic nematic cell-cell stresses. We then analyze integer topological defects that are formed by muscle precursor cells (C2C12 myoblasts) when confined to small circular domains~\cite{PauScience}. Combining our experimental data and our theory allows us to determine material parameters of myoblast monolayers. The experiments analyzed in this work are published in~\cite{PauScience} and part of this work is published in an accompanying letter~$[$Letter$]$.

\section{Hydrodynamic description of monolayers of anisotropic cells }\label{Sec1}

In this section, we develop a phenomenological description of monolayers of elongated cells. After presenting the dynamic equations, we apply them to a monolayer of C2C12 myoblasts confined to a circular domain~\cite{PauScience}.

\subsection{Hydrodynamic fields and conservation equations}

To describe cell monolayers, we use a hydrodynamic approach and start by identifying the hydrodynamic variables characterizing such systems. Let us consider first the two-dimensional cell number density $n$. Cell division and growth occur on a time scale of ten hours. Focussing on shorter time scales, we can neglect these processes and write the conservation equation
\begin{align}
\partial_tn+\partial_{\gamma}(n v_{\gamma})&=0,\label{eq:massbalance}
\end{align}
where $\gamma$ represents the cartesian coordinates in the substrate plane and $\mathbf{v}$ is the in-plane velocity field. We adopt the Einstein convention such that summation over repeated indices is tacitly assumed. In principle, also the chemical fuel, adenosine-triphosphate (ATP), and its hydrolysis products, adenosine-diphosphate (ADP) and inorganic phosphate P$_i$, satisfy conservation equations. However, in our experiments, the cells metabolize nutrients provided by the buffer to replenish consumed ATP from ADP and P$_i$~\cite{PauScience}. Therefore, we assume that the concentrations of ATP, ADP, and P$_i$ are homogenous and constant in time.  

Next, we consider momentum conservation. In our experiments, the Reynolds number $Re$ is small: The C2C12 myoblasts were confined to small circular domains of radius $\sim 100$~$\mu$m and moved at a typical speed $\sim 0.5$~$\mu$m/min. In addition, taking the density of water for the mass density of cells \cite{Grover2011} and using the viscosity of epithelial tissues, which is $\sim 10^9$ times that of water \cite{Blanch-Mercader2017}, we find $Re\sim 10^{-15}-10^{-16}$. We thus consider the overdamped limit and the conservation of momentum is expressed through force balance.

In our experiments, the lateral extension of C2C12 monolayers is an order of magnitude larger than its height, ~50~$\mu$m \textit{vs}~10~$\mu$m. In this limit, a thin-film approximation can be used to turn the 3d force balance equation into an effective 2d description for the height-averaged stress and the height itself~\cite{Kruse2006}. We neglect any fluctuations in the latter and assume it to be uniform, such that force balance is captured by the following effective equation
\begin{eqnarray}
\partial_{\beta}\sigma^\mathrm{tot}_{\alpha\beta}=\xi v_{\alpha}-T_0 p_{\alpha}.\label{eq:forcebalance}
\end{eqnarray}
Here $\sigma^\mathrm{tot}_{\alpha\beta}$ are the cartesian components of the in-plane total mechanical stress tensor obtained after averaging with respect to the height. On the right hand side of the equation, the external force density results from interactions of the cells with the substrate. No net force and torque is applied on the monolayers as a result of these interactions. 

The external force density has two components: $\xi\mathbf{v}$ describes friction between the monolayer and the substrate, whereas $T_0\mathbf{p}$ is the traction force of the cells. The friction force depends on the velocity field $\mathbf{v}$. The traction force is independent of the velocity $\mathbf{v}$. It results, for example, from retrograde cytoskeletal flows in lamellipodia or from stress-fiber contraction transmitted to the substrate via long-lived adhesion points. The direction of the traction force derives  from the local average orientation of these cellular structures, which is captured by the polarization field $\mathbf{p}$.  Fluctuations around the average orientation are accounted for by higher order fields, like the nematic tensor $\mathsf{Q}$ \cite{DeGennes1995}. Here, we assume that such terms are determined by $\mathbf{p}$, for example, $\mathsf{Q}\sim\mathbf{p}\mathbf{p}$. A possible nematic contribution to the traction force will be discussed in Sec.~\ref{sec:ActiveNematicForce}.

\subsection{Constitutive relations}

To close the system of equations describing the dynamics of the myoblast monolayer, expressions for the total stress $\mathsf{\sigma}^\mathrm{tot}$ and the time evolution of the polarization field $\mathbf{p}$ are needed. To obtain such expressions, we follow the standard approach of non-equilibrium thermodynamics~\cite{DeGroot1963}. It consists of first identifying pairs of conjugated thermodynamic forces and fluxes by inspecting the time derivative of the free energy. In a second step, the fluxes are expressed to linear order in terms of the forces, where the coupling coefficients obey the Onsager relations. 

Here, we choose the following quantities as thermodynamic forces~\cite{Kruse2005}: the symmetric part of the velocity gradient tensor with components $v_{\alpha\beta}=(\partial_\alpha v_\beta+\partial_\beta v_\alpha)/2$, the field $\mathbf{h}=-\delta\mathcal{F}/\delta\mathbf{p}$, where  $\mathcal{F}$ is the equilibrium free energy, and the difference between the chemical potentials of ATP, ADP and P$_i$ $\Delta\mu=\mu_\mathrm{ATP}-\mu_\mathrm{ADP}-\mu_\mathrm{P}$. The corresponding thermodynamic fluxes are given by the deviatory stress tensor $\mathsf{\sigma} =\mathsf{\sigma}^\mathrm{tot}-\mathsf{\sigma}^e$, the co-rotational convective derivative of the polarization field $D\mathbf{p}/Dt$, and the rate $r$ of ATP-hydrolysis~\cite{Kruse2005}. As we assume constant densities of ATP, ADP, and P$_i$ we do not consider $r$ any further. The Ericksen stress $\mathsf{\sigma}^e$ is a generalization of the hydrostatic pressure, see App.~\ref{sec:EricksenStress}. In the context of liquid crystals \cite{DeGennes1995}, $\mathbf{h}$ is called the molecular field. It describes the restoring forces associated with deformations of $\mathbf{p}$. The co-rotational convective derivative of the polarization field is given by 
\begin{align}
\frac{D}{Dt}p_\alpha &=\partial_t p_{\alpha}+v_\beta \partial_\beta p_\alpha+\omega_{\alpha\beta}p_\beta.\label{eq:corotCovecDerivative}
\end{align}
Here, $\omega_{\alpha\beta}=(\partial_\alpha v_\beta-\partial_\beta v_\alpha)/2$ is the antisymmetric part of the velocity gradient tensor. 

Before proceeding to discuss the constitutive equations, let us first note that there is some freedom in choosing the stress tensor. Only the divergence of the stress has a physical significance, so one can always add a divergence-free component to the stress tensor. We adopt the same choice as in Ref.~\cite{Joanny2007,Furthauer2012}, such that the components of the antisymmetric part of the deviatory stress are 
\begin{align}
\sigma^a_{\alpha\beta}&=\frac{1}{2}\left(p_\alpha h_\beta-p_\beta h_\alpha\right).  
\end{align}
The symmetric part $\mathsf{\sigma}^s$ of the deviatory stress and the co-rotational convective derivative of the polarization field are obtained, as mentioned above, by expressing these fluxes in terms of the thermodynamic forces in lowest order. Explicitly, we find
\begin{align}
\sigma_{\alpha\beta}^s& = 2\eta\left( v_{\alpha\beta}-\frac{1}{2}v_{\gamma\gamma}\delta_{\alpha\beta}\right)+\bar\eta v_{\gamma\gamma}\delta_{\alpha\beta}+\frac{\nu}{2}\left(p_{\alpha}h_\beta+p_\beta h_{\alpha}-p_\gamma h_\gamma \delta_{\alpha\beta}\right)+\nu' p_\gamma h_\gamma \delta_{\alpha\beta} \nonumber\\
& \quad\quad-\left(p_{\alpha}p_\beta-\frac{1}{2}p_\gamma p_\gamma\delta_{\alpha\beta}\right)\zeta\Delta\mu-\delta_{\alpha\beta}\zeta'\Delta\mu-p_\gamma p_\gamma \delta_{\alpha\beta}\zeta'' \Delta\mu\label{eq:devstresstensor}\\ 
\frac{D}{Dt}p_\alpha&= \frac{h_\alpha}{\gamma}-\nu \left(v_{\alpha\beta}-\frac{1}{2}v_{\gamma\gamma}\delta_{\alpha\beta}\right)p_\beta-\nu' v_{\beta\beta}p_\alpha\label{eq:dinamicadirector}
\end{align}
In the expression for the symmetric part of the deviatory stress $\mathsf{\sigma}^s$, the first two terms account for viscous stresses, where the coefficient $\eta$ and $\bar\eta$, respectively, are the shear and bulk viscosities of the cell monolayer. The following two terms couple the mechanical stress to the field $\mathbf{h}$. All these terms also appear in the stress of liquid crystals~\cite{DeGennes1995}. The remaining terms couple the mechanical stress to ATP-hydrolysis and thus denote the active components of the stress. For our choice of the sign of the stress tensor, positive values of $\zeta$, $\zeta'$, and $\zeta''$ correspond to extensile active stresses. Let us remark that also the expressions for the friction and traction forces in Eq.~\eqref{eq:forcebalance} could be obtained from similar arguments~\cite{Julicher2009}. In this way, the traction force is coupled to ATP-hydrolysis.

In Equation~\eqref{eq:dinamicadirector}, the first term captures relaxation of the polarization field with $\gamma$ being a rotational viscosity. The parameters $\nu$ and $\nu'$ are the so-called flow-alignment parameters. They describe the response of the polarization field to gradients in the velocity field $\mathbf{v}$. In particular, $\nu$ describes the response to shear flows, whereas $\nu'$ that to divergent flows. Note that, in this equation, we have omitted an active term, that is a coupling to $\Delta\mu$. Such a term would be of the form $p_\alpha\lambda\Delta\mu$. We will see in Sect.~\ref{sec:activeAlignment} that this amounts to a renormalization of parameters.   

Explicit expressions for the Ericksen stress $\mathsf{\sigma}^e$ and the field $\mathbf{h}$ are obtained by fixing the equilibrium free energy  $\mathcal{F}$ of the system. We choose
\begin{align}
\mathcal{F}&=\int_\mathcal{A}\left\{\frac{B}{2}\left(1-\frac{n}{n_0}\right)^2+\frac{\chi}{2}p_{\alpha}^2+\frac{{\cal K}}{2}(\partial_{\alpha}p_{\beta})^2\right\}da. \label{eq:freeenergy} 
\end{align}
The first term penalizes deviations of the cell density from the reference density $n_0$, where $B$ is the corresponding bulk modulus. The remaining terms capture the elastic energy associated with distortions of the polarization field similar to the free energy used for liquid crystals~\cite{DeGennes1995}. As suggested by our experiments, see Sect.~\ref{sec:experiments} below, we consider  $\chi>0$ meaning that the preferred bulk equilibrium state is disordered. The energy cost associated with gradients of the polarization field is accounted for by the final term. It is equal to the Frank energy in the one-constant approximation with modulus $\mathcal{K}$. This approximation is appropriate for the experimental system as we show in Sec.~\ref{sec:FrankConstants}.  

Let us remark that the term of uniform isotropic active stress $\zeta'\Delta\mu\mathbb{I}$ in Eq.~\ref{eq:devstresstensor} amounts to a renormalization of parameters. Explicitly, the bulk modulus $B$ and the reference density $n_0$ are transformed as follows: $B\rightarrow B-2\zeta'\Delta\mu$ and $n_0\rightarrow n_0\sqrt{1-2\zeta'\Delta\mu/B}$. For large enough positive $\zeta'\Delta\mu$, the effective bulk modulus $B$ is negative, which may lead to mechanical instabilities that are similar to those found in other contexts \cite{Joanny}. Henceforth, we consider $\zeta'\Delta\mu=0$ and exclude this scenario as we have not found signatures of such instabilities in our experiments.

Let us briefly summarize the parameters appearing in our description. Active processes  are captured by the magnitude of the traction force $T_0$ and the parameters $\zeta$ and $\zeta'$ coupling ATP hydrolysis to the mechanical stress. Dissipation occurs through rearrangements of the polarization, the viscous dissipation, and friction with the substrate, which are, respectively, controlled by the coefficients $\gamma$, $\eta$, $\bar\eta$, and $\xi$.  Flow alignement of the polarization is governed by $\nu$ and $\nu'$ and, finally, there are three elastic moduli, namely, $B$, $\chi$, and $\mathcal{K}$.

\subsection{Myoblast monolayers}
\label{sec:experiments}

We studied the collective behavior of C2C12 cells confined to fibronectin-coated circular domains with radii between $50$~$\mu$m and $150$~$\mu$m. In the following, we describe the main features of the methods used. For further experimental details, see~\cite{PauScience}. 

Individually, C2C12 mouse myoblasts move at speeds of $20-50$~$\mu$m/h, and they can assume an elongated shape around $50$~$\mu$m in length and $10$~$\mu$m in width \cite{Sheets2013}. Extended C2C12 myoblast monolayers spontaneously generate long range nematic  order~\cite{Duclos2017,Kawaguchi2017,PauScience}. This corresponds to $\chi<0$ in the equilibrium free energy~\eqref{eq:freeenergy}. Correspondingly, these monolayers can present half-integer topological defects~\cite{Kawaguchi2017}.

\begin{figure}[t] %  figure placement: here, top, bottom, or page
\centering
\includegraphics[width=0.45\textwidth]{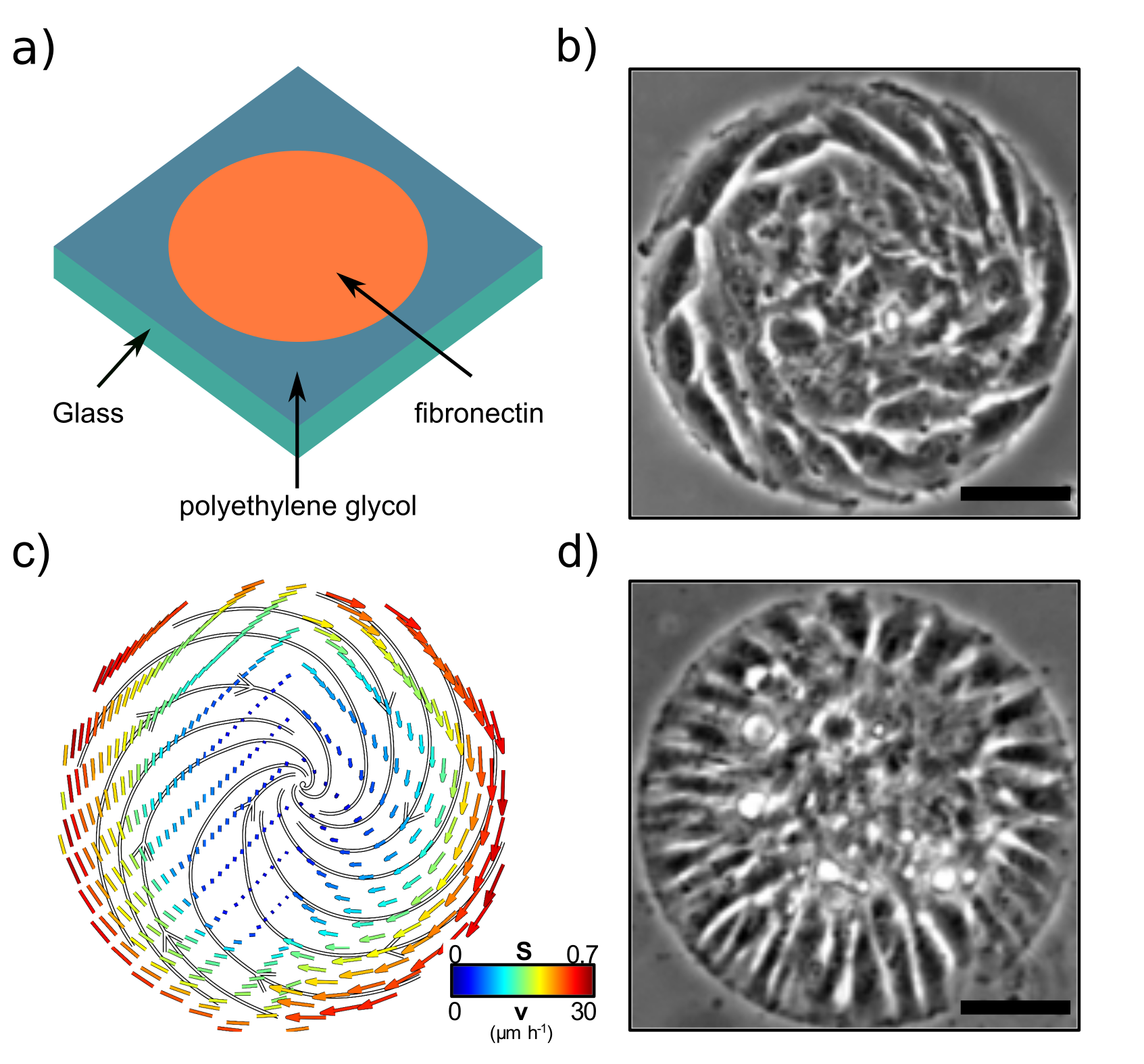}
\caption{(online color) Confined C2C12 monolayers. a) Schematic of the experimental setup. b) Phase-contract image of a spiral in a circular domain of 100~$\mu$m radius. c) Orientational order (left) and velocity fields (right) averaged over $N=12$ spirals. Colors correspond to $S$ and speeds, see legend. Gray lines: velocity stream lines. d) Phase-contrast image of an aster in a circular domain of 100~$\mu$m radius. Scale bar in (b,d): 50~$\mu$m.}\label{fig001}
\end{figure}

In our experiments, cells were confined to fibronectin-coated circular domains by coating the surrounding with non-adhesive polyethylene glycol, Fig.~\ref{fig001}a.  Over the course of our experiments, the cell number increases by proliferation. After a transient, cells formed a uniform monolayer without visible cell-free gaps. In contrast to extended monolayers, in our small islands, we observe polar order near the domain boundary as reflected by continuous lamellipodial activity. Correspondingly, the cell monolayers arranged into integer topological defects with a disorganized center. We thus chose polar traction forces and $\chi>0$ in the free energy~\eqref{eq:freeenergy}. 

At low densities, we found that cell monolayers spontaneously arranged into spirals that collectively rotated, Fig.~\ref{fig001}b. The orientation of the cell bodies at the interface of the circular domains was approximately tangential, and the average rotational speed was on the order of $30$~$\mu$m/h, Fig.~\ref{fig001}c. As the cell number increased further, we found that cells at the periphery changed their orientation by aligning their bodies perpendicularly to the circular interface thus forming an aster, see Fig.~\ref{fig001}d. In this case, the collective rotation was lost. Further evolution of these cell monolayers led to 3d multicellular protrusions featuring long-range nematic order and collective cell dynamics perpendicular to the confinement plane, see \cite{PauScience}.

From phase-contrast movies, particle velocimetry techniques were used to determine a coarse-grained velocity field. From the same movies, we determined a coarse-grained orientational field via a structure tensor method \cite{Puspoki2016}. For a given 2d intensity pattern, this technique computes the direction of the minimal and maximal intensity anisotropy as the eigenvectors of a 2d structure matrix obtained from intensity gradients. Then, we set the orientational field parallel to the eigenvector with minimal eigenvalue. A representative example of both time-averaged fields for spiral configurations is shown in Fig.~\ref{fig001}c.

\subsection{Circular confinement}

In the following, we apply the equations derived in the previous sections to cell monolayers confined to circular islands. We therefore express the equations in polar coordinates $r$ and $\theta$. We  focus on steady state solutions and assume that they are invariant with respect to rotations around the center of the island. Finally, we determine the boundary conditions for this situation. 

\subsubsection{Steady state equations in polar coordinates}

We start with the conservation equation \eqref{eq:massbalance} for the cell number density. In steady state and assuming rotational invariance, it becomes
\begin{align}
\partial_{r}(n v_{r})+\frac{n v_r}{r}&=0\label{eq:massbalanceprimer}.
\end{align}
As will be detailed below, there are no flows across the domain boundaries, such that $v_r=0$ in steady state.

For the polarization field $\mathbf{p}$, we introduce the magnitude or 'polar order parameter' $S$ and the angle $\psi$ with respect to the radial direction, such that $p_r=S \cos(\psi)$ and $p_\theta=S \sin(\psi)$. In terms of the variables $S$ and $\psi$, the dynamic equation~\eqref{eq:dinamicadirector} for the polarization field reads
\begin{align}
\frac{h_\parallel}{\gamma}-\nu S v_{r\theta}\sin(2\psi)&=0\label{eq:dinamicadirectorparticular1}\\
\frac{h_\perp}{\gamma}+S v_{r\theta}\left(1-\nu \cos(2\psi)\right)&=0 .\label{eq:dinamicadirectorparticular2}
\end{align}
In these expressions, $h_\parallel$ and $h_\perp$ are the components of the field $\mathbf{h}$ parallel and perpendicular to $\mathbf{p}$. The explicit expressions of $h_\parallel$ and $h_\perp$ are given in Eqs.~\eqref{eq:molecularfieldpara} and \eqref{eq:molecularfieldperp} in App.~\ref{sec:molecularfield}. Furthermore, $v_{r\theta}=(\partial_r v_\theta-v_\theta/r)/2$ is the off-diagonal component of the symmetric part of the velocity gradient tensor. The components  $v_{rr}$ and $v_{\theta\theta}$ vanish at steady state.

Using the variables $S$ and $\psi$, the components of the deviatory stress can  be written as
\begin{align}
\sigma_{rr,\theta\theta}&=\mp\frac{1}{2}S^2 \cos(2\psi)\zeta\Delta\mu-S^2\zeta''\Delta\mu  \nonumber\\
&\pm\frac{\nu}{2}S\left(h_\parallel \cos(2\psi)-h_\perp \sin(2\psi)\right)   +\nu' S h_\parallel \label{eq:devStressTensorRR} \\
\sigma_{r\theta,\theta r}&=2\eta v_{r\theta}-\frac{1}{2}S^2 \sin(2\psi)\zeta\Delta\mu  \nonumber\\
&+\frac{\nu}{2}S\left(h_\parallel \sin(2\psi)+h_\perp \cos(2\psi)\right) \pm\frac{S h_\perp}{2},\label{eq:devStressTensorRTheta} 
\end{align}
where the upper (lower) signs correspond to the first (second) index pair. 
The force balance equation \eqref{eq:forcebalance} takes the form
\begin{align}
\partial_{r}\sigma_{rr}^\mathrm{tot}+\frac{\sigma_{rr}^\mathrm{tot}-\sigma_{\theta\theta}^\mathrm{tot}}{r}&=-T_0 S \cos(\psi)\label{eq:forcebalance1}\\
\partial_{r}\sigma_{\theta r}^\mathrm{tot}+\frac{\sigma_{\theta r}^\mathrm{tot}+\sigma_{r\theta}^\mathrm{tot}}{r}&=\xi v_{\theta}-T_0 S \sin(\psi).\label{eq:forcebalance2}
\end{align}
By employing the Gibbs-Duhem relation \eqref{eq:FBStressEricksenParB}, we can furthermore eliminate the Ericksen stress in Eq.~\eqref{eq:forcebalance2} and obtain
\begin{align}
\partial_{r}\sigma_{\theta r}+\frac{2\sigma_{\theta r}}{r}&=\xi v_{\theta}-T_0 S \sin(\psi).\label{eq:forcebalance3}
\end{align}

\subsubsection{Boundary conditions}

It remains to fix the conditions on the fields at the boundary of the island at $r=R$, where $R$ is the radius of the domain. Compatible with our experiments, we impose that the there is no flux of material into the domain at the boundary. At the same time, there is no tangential force applied to the cell monolayer at the edge of the domain. For the boundary conditions on the polarization field, let us first note that the polar order parameter is maximal at the boundary. Without loss of generality, we fix this value to be one. Furthermore we impose that there are no gradients in $\psi$ at the boundary. In summary, we thus have 
\begin{align}
S|_{r=R}&=1\label{eq:boundarycon1}\\
\partial_r\psi|_{r=R}&=0\label{eq:boundarycon2}\\
\sigma_{\theta r}^\mathrm{tot}|_{r=R}&=0\label{eq:boundarycon3}\\
v_r|_{r=R}&=0. \label{eq:boundarycon4}
\end{align}
Note that the total cell number is conserved and thus a parameter of our system. 

In our experiments, the monolayers are disordered in the center of the domains, and we impose $S=0$ at $r=0$. Due to our assumption of rotational invariance, we also need to impose regularity of the solutions at $r=0$. In total we have
\begin{align}
S|_{r=0}&=0\label{eq:boundarycon5}\\
\partial_r\psi|_{r=0}&=0\label{eq:boundarycon6}\\
v_{\theta}|_{r=0}&=0\label{eq:boundarycon7}\\
v_{r}|_{r=0}&=0.\label{eq:boundarycon8}
\end{align}

\section{Active forces in integer topological defects}\label{sec:Activeforces}

Materials with orientational order are prone to exhibit singularities in the corresponding order parameter. Such singularities are called topological defects. They are characterized by their 'charge', that is, the number of turns of the polarization vector upon moving it along a closed path around the singularity. The most common types are defects with charges $\pm$1/2 and $\pm1$. 

As mentioned in the Introduction, topological defects have been related to biological processes in cell monolayers~\cite{Saw2017,Kawaguchi2017,Maroudas-Sacks2020,PauScience}. For a better understanding of the mechanics of defects in monolayers under confinement, we analyze now the active force density associated with +1 defects. In our description, activity enters in different terms, namely, in the traction force $T_0\mathbf{p}$ and in the stress via 
\begin{align}
\sigma^\mathrm{act}_{\alpha\beta}&=-\left(p_{\alpha}p_\beta-\frac{1}{2}p_\gamma p_\gamma\delta_{\alpha\beta}\right)\zeta\Delta\mu- p_\gamma p_\gamma\delta_{\alpha\beta}\zeta''\Delta\mu.
\end{align}
The surface active force density then is 
\begin{align}
\mathbf{f}^{a,s}&=T_0 \mathbf{p}+\nabla\cdot\mathsf{\sigma}^\mathrm{act}.\label{eq:App5}
\end{align}
In addition, there is a line active force density at the boundary of the circular domain with radius $R$
\begin{align}
\mathbf{f}^{a,l}&=-\mathsf{\sigma}^\mathrm{act}\cdot\hat{\mathbf{r}}|_{r=R},\label{eq:App8}
\end{align}
where $\hat{\mathbf{r}}$ is the radial unit vector.

\begin{figure}[b] %  figure placement: here, top, bottom, or page
	\centering
	\includegraphics[]{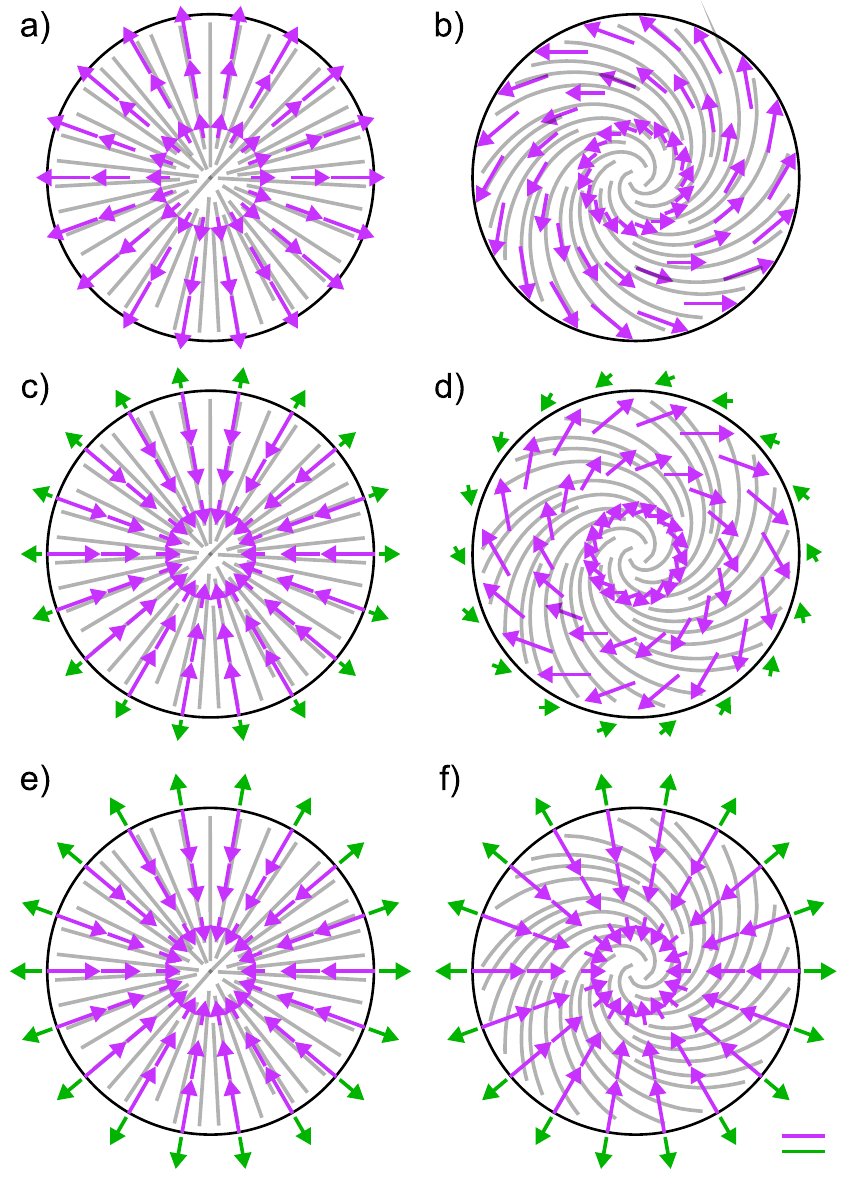}
	\caption{(online color) Active forces associated with integer topological defects: asters (a,c,e), and spirals (b,d,f). Active forces only generated by traction forces $T_0\mathbf{p}$ (a,b), by anisotropic active stresses proportional to $\zeta\Delta\mu$ (c,d), and by isotropic active stresses proportional to $\zeta''\Delta\mu$ (e,f). Gray lines indicate the polarization field, which points outwards. The angle of the spiral is $\psi_0=\pi/3$ (b,d,f). Magenta arrows: surface active force density at $r/R=\{1/3, 2/3, 1\}$, $\mathbf{f}^{a,s}$ in Eq.~\eqref{eq:App7a}. Green arrows: line active force density, $\mathbf{f}^{a,l}$ in Eq.~\eqref{eq:App7b}. Black circle: boundary at $r=R$. The shafts of the magenta arrows are scaled by $\mathbf{f}^{a,s}(r=R)$ and of the green arrows by $R\mathbf{f}^{a,s}(r=R)$. Scale bars indicate $\mathbf{f}^{a,s}(r=R)=R\mathbf{f}^{a,s}(r=R)=1$. We assumed $T_0,\zeta\Delta\mu,\zeta''\Delta\mu>0$.}
	\label{fig0}
\end{figure}

The simplest form of +1 defects corresponds to spirals with constant angle $\psi=\psi_0$. In the cases, $\psi_0=0$, $\pi$ and $\psi_0=\pm\pi/2$, the spirals turn into asters or vortices, respectively. For the polar order parameter $S$, we will assume a linear dependence on the radial coordinate $r$, such that $S=r/R$. As we will see below, this is a solution to our equations in the limit of small radius $R$. Using expressions \eqref{eq:devStressTensorRR}-\eqref{eq:devStressTensorRTheta} for the components of the active stress tensor, we obtain
\begin{align}
\mathbf{f}^{a,s}&=\left(T_0 R \cos{(\psi_0)}-2\zeta\Delta\mu  \cos{(2\psi_0)}-2\zeta''\Delta\mu \right)\frac{r \hat{\mathbf{r}}}{R^2}\nonumber\\
&\quad+\left(T_0 R \sin{(\psi_0)}-2\zeta\Delta\mu  \sin{(2\psi_0)} \right)\frac{r \hat{\bm{\theta}}}{R^2},\label{eq:App7a}\\
\intertext{and}
\mathbf{f}^{a,l}&=\left(\frac{\zeta\Delta\mu}{2}\cos(2\psi_0)+\zeta''\Delta\mu\right)\hat{\mathbf{r}}+\left(\frac{\zeta\Delta\mu}{2} \sin(2\psi_0)\right) \hat{\bm{\theta}}\label{eq:App7b}
\end{align}
where $\hat{\bm{\theta}}$ is the azimuthal unit vector. Figure~\ref{fig0} presents these force densities for asters and spirals.

For asters with $\psi_0=0$ both, the surface and the line active force densities only have radial components, see Fig.~\ref{fig0}a,c,e. In this case, $\mathbf{f}^{a,s}$ is pointing towards the center if $T_0 R-2(\zeta+\zeta'')\Delta\mu<0$ and vice versa. 

For spirals, the surface and the line active force density has a radial and an azimuthal component, see Fig.~\ref{fig0}b,d,f. For spirals with $\psi_0>\pi/4$ but otherwise the same parameter values as for asters, the radial component of $\mathbf{f}^{a,s}$ can point away from the center, Eq.~\eqref{eq:App7a}. The same effect can be observed for the radial component of $\mathbf{f}^{a,l}$, Eq.~\eqref{eq:App7b}. The azimuthal components of $\mathbf{f}^{a,s}$ and $\mathbf{f}^{a,l}$ are independent of the isotropic active stress proportional to $\zeta''\Delta\mu$, Eqs.~\eqref{eq:App7a} and \eqref{eq:App7b}.

For vortices with $\psi_0=\pi/2$, the traction forces generate an azimuthal component in the surface active force density. In this case, $\mathbf{f}^{a,s}$ is pointing towards the center if $2(\zeta-\zeta'')\Delta\mu<0$ and vice versa.

In the following two sections, we discuss in detail the steady states of integer topological defects.

\section{Asters} %polar character is a pure boundary effect
\label{sec:asters}

We consider first the special case of an aster, where $\psi_0=0$. In that case, the azimuthal velocity $v_\theta$ vanishes by symmetry. Equation~\eqref{eq:dinamicadirectorparticular2} then implies $h_\perp=0$, showing that the aster is a solution of our system. It follows from Equation~\eqref{eq:dinamicadirectorparticular1} that also $h_\parallel=0$. Using this result in Equation~\eqref{eq:molecularfieldpara} and the boundary conditions~\eqref{eq:boundarycon1} and \eqref{eq:boundarycon5}, the polar order parameter $S$ can be calculated. The general solution is given by a Bessel function. Since in our experiments, we see a single defect per island~\cite{PauScience}, we focus on the limit $R^2\ll {\cal K}/\chi$. In that case, the penetration length of the boundary polar order $\sqrt{{\cal K}/\chi}$ is larger than the system size $R$ and $S=r/R$. For larger island radii, multiple defects were reported for C2C12 monolayers~\cite{Duclos2017}.

\begin{figure}[b] %  figure placement: here, top, bottom, or page
	\centering
	\includegraphics[]{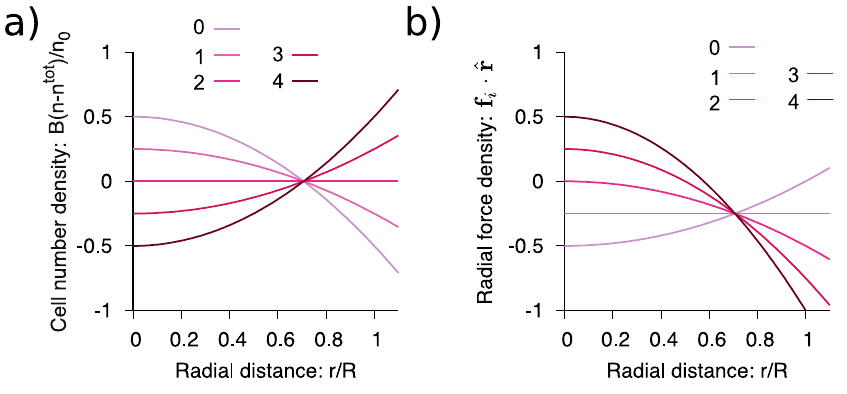}
	\caption{(online color) Steady state profiles for asters. a) Cell number density $B(n-n^\mathrm{tot})/n_0$, Eq.~\eqref{eq:densityasters2}, b) radial force density $\mathbf{f}_i\cdot\hat{\mathbf{r}}$, Eq.~\eqref{eq:forceinneraster}, as a function of the radial distance $r$ for varying values of the dimensionless ratio $T_0 R/\zeta\Delta\mu$ as indicated in the legend. We consider $\zeta''\Delta\mu=0$ (a) and $-\frac{\zeta''\Delta\mu}{2}-B\frac{n^\mathrm{tot}-n_0}{n_0}=0$ (b). Units are set by $\zeta\Delta\mu=R=1$.}
	\label{fig00}
\end{figure}

It remains to determine the cell number density for the aster. To this end, we employ the radial component of the force balance Eq.~\eqref{eq:forcebalance1}. Note that the azimuthal component, Eq.~\eqref{eq:forcebalance2}, is automatically satisfied by symmetry. In the limit $R^2\ll {\cal K}/\chi$, the non-vanishing components of the total stress tensor read
\begin{align}
\sigma_{rr}^\mathrm{tot}&=\frac{B}{2}\left(1-\frac{n^2}{n_0^2}\right)-\left(\frac{1}{2}\zeta\Delta\mu +\zeta''\Delta\mu\right)\frac{r^2}{R^2} \\ 
 \sigma_{\theta\theta}^\mathrm{tot}&=\frac{B}{2}\left(1-\frac{n^2}{n_0^2}\right)+\left(\frac{1}{2}\zeta\Delta\mu - \zeta''\Delta\mu\right)\frac{r^2}{R^2}. 
\end{align}
In the limit that there are only small deviations from the reference density $n_0$, the solution to Eq.~\eqref{eq:forcebalance1} is
\begin{align}
\frac{n-n_0}{n_0}&\approx\frac{1}{B}\left[\left(\frac{R}{2}T_0-\zeta \Delta\mu-\zeta''\Delta\mu\right)\frac{r^2}{R^2}+n_c \right], \label{eq:densityasters}
\end{align}
where $n_c$ is an integration constant. If the total cell number in the circular island is $n^\mathrm{tot}\pi R^2$, then
\begin{align}
\frac{n-n^\mathrm{tot}}{n_0}&\approx\frac{1}{B}\left(\frac{R}{2}T_0-\zeta\Delta\mu -\zeta''\Delta\mu\right)\left(\frac{r^2}{R^2}-\frac{1}{2}\right). \label{eq:densityasters2}
\end{align}
In Figure~\ref{fig00}a, we show the density as a function of the radial coordinate for different ratios $T_0R/\zeta\Delta\mu$ and fixed $\zeta''\Delta\mu$.

Next, let us determine the momentum that the monolayer in the aster configuration exchanges with the environment. As the velocity $\mathbf{v}=0$, the force exerted by the monolayer on the substrate is
\begin{align}
\mathbf{t}=-T_0\frac{r}{R}\hat{\mathbf{r}}.\label{eq:Taster}
\end{align} 
At the confinement boundary $r=R$ and to first order in $n^\mathrm{tot}/n_0$, the local force density per unit length is
\begin{align}
\mathbf{f}_o &= -\mathsf{\sigma}^\mathrm{tot}(r=R)\cdot\hat{\mathbf{r}}\\
&= \left(\frac{T_0 R}{4}+\frac{\zeta''\Delta\mu}{2}+B\frac{n^\mathrm{tot}-n_0}{n_0}\right)\hat{\mathbf{r}}. \label{eq:forceouteraster}
\end{align}
From Eqs.~\eqref{eq:Taster} and \eqref{eq:forceouteraster}, we see that the total force on the monolayer 
\begin{align}
\mathbf{F}^\mathrm{tot}&=\int_\mathcal{A} \mathbf{t} da+\int_{\partial\mathcal{A}}\mathbf{f}_o dl\label{eq:Ftotal}
\end{align}
vanishes, $\mathbf{F}^\mathrm{tot}=\mathbf{0}$. Because the forces are all radial, also the total torque 
\begin{align}
\mathbf{M}^\mathrm{tot}&=\int_\mathcal{A} \mathbf{r}\times\mathbf{t} da+\int_{\partial\mathcal{A}}R\hat{\mathbf{r}}\times\mathbf{f}_o dl\label{eq:Mtotal}
\end{align}
is zero. Therefore, neither a net force nor a net torque results from interactions between the monolayer and the substrate in steady state asters.  

In our experiments~\cite{PauScience}, we used circular elastic pillars placed in the center of the circular domain to measure the force exerted by the monolayer. Neglecting deviations from the profiles calculated above that are caused by the finite diameter of the pillar, this force is
\begin{align}
\mathbf{f}_i&=\mathsf{\sigma}^\mathrm{tot}(r)\cdot\hat{\mathbf{r}}\\
&=\left[\frac{R}{2}\left(\frac{1}{2}-\frac{r^2}{R^2}\right)T_0 +\frac{1}{2}\left(\frac{r^2}{R^2}-1\right)\zeta\Delta\mu-\frac{1}{2}\zeta''\Delta\mu-B\frac{n^\mathrm{tot}-n_0}{n_0}\right]\hat{\mathbf{r}},\label{eq:forceinneraster}
\end{align}
see Fig.~\ref{fig00}b. Although this expression is correct only in the limit, where the diameter of the pillars tends to zero, it gives an approximate value for pillars with finite diameter.  

\section{Spirals}
\label{sec:spiral}

\begin{figure}[b] %  figure placement: here, top, bottom, or page
 	\centering
 	\includegraphics[]{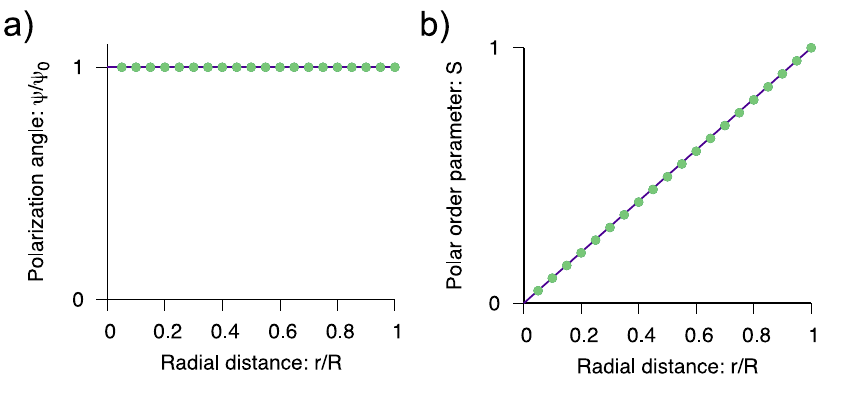}
 	\caption{(online color) Steady-state profiles of the orientational order in spirals and with $R^2\ll{\cal K}/\chi$. a) Polarization angle $\psi$ and b) polar order parameter $S$. Purple lines: $S=r/R$ and $\psi=\psi_0$, respectively. Green dots: numerical solution of the dynamic equations. Parameter values are  $\chi=0.1$, $\nu=-1.4$, $\zeta=10^{-2}$, $T_0=0$, $\eta=10^2$, and $\xi=1$ with the units being set by $R={\cal K}=\gamma=1$. For these parameter values $|\gamma\nu v_{r\theta}\sin(2\psi_0)|<2*10^{-5}\ll \chi$.}\label{fig3}
 \end{figure}

In the following, we turn to the case of a general topological defect with charge +1, where $\psi(r)$ takes on an arbitrary constant value $\psi_0$. A constant value of $\psi$ implies $h_\perp=0$, see Eq.~\eqref{eq:molecularfieldperp}. Its value is fixed by the steady state Eq.~\eqref{eq:dinamicadirectorparticular2}, which implies $\nu \cos(2\psi_0)=1$. This condition requires $|\nu|\geq1$ for a real solution $\psi_0$. Note that $\psi(r)=\psi_0$ also satisfies the boundary conditions~\eqref{eq:boundarycon2} and \eqref{eq:boundarycon6}, see Fig.~\ref{fig3}a for a comparison of the analytic result with a numeric solution of the dynamic equations. Without restriction of generality we consider $0<\psi_0<\pi/2$.

Next, we consider Eq.~\eqref{eq:dinamicadirectorparticular1} with $h_\parallel$ given by Eq.~\eqref{eq:molecularfieldpara}. As for the case of asters discussed above, we focus on the case $R^2\ll\mathcal{K}/\chi$. Furthermore, we  consider that $|\gamma\nu v_{r\theta}\sin(2\psi_0)|\ll \chi$. In this limit, flow alignment does not lead spontaneously to orientational order and the solution to Eq.~\eqref{eq:molecularfieldpara} is $S=r/R$, see Fig.~\ref{fig3}b. 

\subsection{Velocity field}

Having obtained the polarization field, we now determine the velocity field. To this end, let us first consider force balance in the azimuthal direction, see Eq.~\eqref{eq:forcebalance3}. Using the expressions for $S$ and $\psi$, we obtain a differential equation for the azimuthal component $v_\theta$ of the velocity
\begin{align}
\partial_{r}\sigma_{\theta r}+\frac{2\sigma_{\theta r}}{r}&=\xi v_{\theta}-T_0\frac{r}{R}\sin(\psi_0),\label{eq:forcebalanceanalytic}
\end{align}
where the off-diagonal component $\sigma_{\theta r}$ of the deviatory stress tensor reads
\begin{align}
\sigma_{\theta r}&=\left(2\eta+\gamma\frac{ r^2 }{2 R^2}\tan(2\psi_0)^2\right) v_{r\theta}-\frac{r^2}{2 R^2} \sin(2\psi_0)\zeta \Delta\mu,\label{eq:devStressTR}
\end{align}
see Eq.~\eqref{eq:devStressTensorRTheta}. The boundary conditions are given by Eqs.~\eqref{eq:boundarycon3} and \eqref{eq:boundarycon7}.

In our system, azimuthal flows are generated by two different active processes, namely, gradients in the active stress, which is proportional to $\zeta\Delta\mu$, and traction forces, which are proportional to $T_0$ as discussed in Sect.~\ref{sec:Activeforces}. Since Eq.~\eqref{eq:forcebalanceanalytic} is linear in $v_\theta$, we discuss these two origins of flows by solving Eq.~\eqref{eq:forcebalanceanalytic} in various limiting regimes that differ in the dominant dissipative mechanism. Explicitly,
\begin{itemize}
\item Regime I, where dissipation is dominated by shear viscosity: $\gamma\tan(2\psi_0)^2\ll \eta$ and $\xi R^2\ll \eta$;
\item Regime II, where dissipation is dominated by relaxation of the polarization field: $\eta\ll \gamma\tan(2\psi_0)^2$ and $\xi R^2\ll \gamma\tan(2\psi_0)^2$; 
\item Regime III, where dissipation is dominated by friction forces with the underlying substrate: $\gamma\tan(2\psi_0)^2 \ll \xi R^2$ and $\eta \ll \xi R^2$.
\end{itemize}
In Regime III we further distinguish the cases $\gamma\tan(2\psi_0)^2\ll\eta$ and $\eta\ll\gamma\tan(2\psi_0)^2$. Whereas in Regimes I and II there are long-ranged flows due to viscous coupling of different parts of the system, in Regime III, flows can be screened beyond distances of the order of the 'friction length' $\ell$, where 
\begin{align}
\ell^2 &= \frac{1}{4\xi}\left(4\eta+\gamma\tan(2\psi_0)^2\right).\label{eq:frictionLength}
\end{align}

\subsubsection{Flows driven by traction forces}

\begin{figure}[t] %  figure placement: here, top, bottom, or page
 	\centering
 	\includegraphics{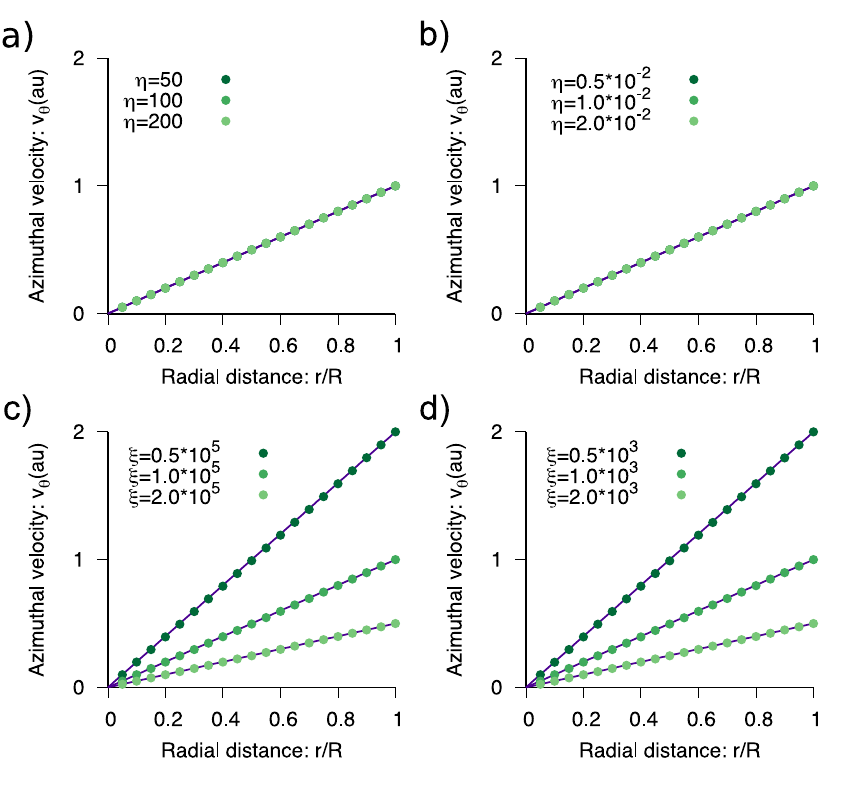}
 	\caption{(online color) Steady-state azimuthal velocity for flows driven by traction forces and with $R^2\ll{\cal K}/\chi$. a) Regime I with $\eta=50$, $100$, $200$ and $\xi=1$, b) Regime II with $100\eta=0.5$, $1$, $2$ and $\xi=10^{-2}$, c) Regime III with $\eta=100$ and $10^{-5}\xi=0.5$, $1$, $2$, and d) Regime III with $\eta=0.01$ and $10^{-3}\xi=0.5$, $1$, $2$. Purple lines: Eq.~\eqref{eq:solution1}. Green dots: numerical solutions of the dynamic equations. Other parameter values are $\chi=10^{-1}$, $\nu=-1.4$,  $T_0=10^{-2}$, and $\zeta\Delta\mu=0$ with the units being set by $R={\cal K}=\gamma=1$.}
 	\label{fig1}
 \end{figure}

In presence of traction forces only, the angular velocity takes the form
\begin{align}
v_{\theta}&=\frac{T_0}{\xi}\frac{r}{R}\sin(\psi_0) .\label{eq:solution1}
\end{align}
As a consequence, the system rotates as a block and no shear flows exist, i.e., $v_{\theta r}=0$. Consequently, neither viscous nor rotational dissipation affects these flows. We have verified numerically that this solution is a good approximation of the flow in Regimes I-III, see Fig.~\ref{fig1}. 

\subsubsection{Flows driven by gradients in active stresses}

In contrast to traction-force driven flows, those driven by gradients in anisotropic active stresses depend on the dominant mechanism of dissipation. We now take $T_0=0$ and consider the different regimes in turn.

For Regimes I and II, the friction term in Eq.~\eqref{eq:forcebalanceanalytic} can be neglected and we have
\begin{align}
\partial_{r}\sigma_{\theta r}+\frac{2\sigma_{\theta r}}{r}&=0.
\end{align}
We thus have $\sigma_{\theta r}=C/r^2$ for some constant $C$. Since $\sigma_{\theta r}$ is finite at $r=0$, it follows that $C=0$. Because the corresponding component of the Ericksen stress also vanishes, $\sigma^e_{\theta r}=0$, see Eq.~\eqref{eq:StressEricksenParrt}, the boundary condition \eqref{eq:boundarycon3} is satisfied. Using Eq.~\eqref{eq:devStressTR}, we can solve $\sigma_{\theta r}=0$ for $v_{\theta r}$ and find that the azimuthal velocity $v_\theta$ is  determined by 
\begin{align}
\frac{1}{2}\left(\partial_rv_\theta-\frac{v_\theta}{r}\right)&=\frac{ r^2\sin(2\psi_0) \zeta \Delta\mu}{4\eta R^2+\gamma r^2 \tan(2\psi_0)^2}.\label{eq:strainratespiral}
\end{align}

In Regime I, the term proportional to $\gamma$ in Eq.~\eqref{eq:strainratespiral} can be neglected and we obtain 
\begin{align}
v_{\theta}&=\frac{ \sin(2\psi_0)\zeta\Delta\mu }{4\eta R^2}r^3+D_\eta r,\label{eq:aux1}
\end{align}
where $D_\eta$ is a constant of integration. Similarly, in Regime II, the term proportional to $\eta$ in Eq.~\eqref{eq:strainratespiral} can be neglected and 
\begin{align}
v_{\theta}=\frac{2\cos(2\psi_0)\zeta\Delta\mu }{\gamma \tan(2\psi_0)}r\ln{(r)}+D_\gamma r,\label{eq:aux2}
\end{align}
where $D_\gamma$ is a constant of integration. Note that both solutions respect the condition $v_\theta=0$ at $r=0$. 

\begin{figure}[b] %  figure placement: here, top, bottom, or page
 	\centering
 	\includegraphics{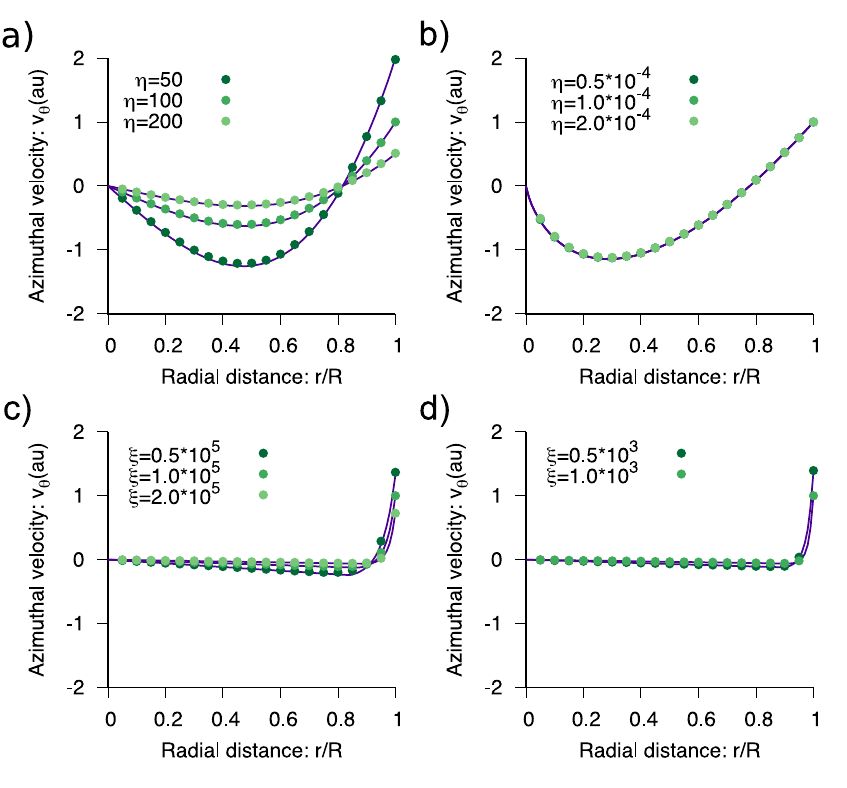}
 	\caption{(online color) Steady-state azimuthal velocity for flows driven by gradients in active stresses and with $R^2\ll{\cal K}/\chi$. a) Regime I with $\eta=50$, $100$, $200$ and $\xi=1$, b) Regime II with $10^4\eta=0.5$, $1$, $2$ and $\xi=10^{-2}$, c) Regime III with $\eta=100$ and $10^{-5}\xi=0.5$, $1$, $2$, and d) Regime III with $\eta=0.01$ and $10^{-3}\xi=0.5$, $1$, $2$. Purple lines: (a) Eq.~\eqref{eq:solvel1}, (b) Eq.~\eqref{eq:solvel2}, (c,d) Eqs.~\eqref{eq:solvel3} and \eqref{eq:solvel4}. Green dots: numerical solution of the dynamic equations. Other parameter values are $\chi=10^{-1}$, $\nu=-1.4$,  $T_0=0$, and $\zeta\Delta\mu=10^{-2}$ with the units being set by $R={\cal K}=\gamma=1$.
}\label{fig2}
 \end{figure}

For vanishing friction, $\xi=0$, the integration constants $D_\eta$ and $D_\gamma$ remain  undetermined. By inserting the solutions \eqref{eq:aux1} and \eqref{eq:aux2} into the force balance Eq.~\eqref{eq:forcebalanceanalytic} and with the friction coefficient $\xi$ being small leads to the respective particular solutions
\begin{align}
v_{\theta}&=\frac{ \sin(2\psi_0)\zeta\Delta\mu}{4\eta}r\left(\frac{r^2}{R^2}-\frac{2}{3}\right)\label{eq:solvel1}\\
\intertext{in Regime I and}
v_{\theta}&=\frac{2\cos(2\psi_0)\zeta\Delta\mu}{\gamma \tan(2\psi_0)} r\ln{(r e^{1/4}/R)}\label{eq:solvel2}
\end{align}
in Regime II. Note that in both cases the azimuthal flow near the outer boundary of the circular domain is opposite to the flow close to the center. The distance from the center at which the flow changes sign is independent of the friction coefficient $\xi$. The stagnation point at which $v_\theta=0$ is placed such that the total torque vanishes, see Sect.~\ref{sec:forceDesnities}. Both solutions agree well with numerical solutions obtained in Regime I and II, see Fig.~\ref{fig2}a,b. 

Let us now turn to Regime III. There, the viscous part of the stress tensor is negligible except in a boundary layer of size $\ell$ that are determined below. Neglecting the viscous stress, the force balance equation~\eqref{eq:forcebalanceanalytic} reads
\begin{align}
-\frac{2 r}{R^2} \sin(2\psi_0)\zeta\Delta\mu&=\xi v_\theta\label{eq:solvel3}
\end{align}
and thus explicitly gives the azimuthal velocity. In the boundary layer, we introduce a new spatial variable $x=(R-r)/R$ and velocity $\tilde v_\theta(x)=v_\theta(R(1-x))$ with $0\le x\ll1$. We then express the force balance equation~\eqref{eq:forcebalanceanalytic} in terms of these variables and keep only terms of order 0 in $x$. Since $\partial_x\tilde v_\theta\sim\tilde v_\theta/(\ell/R)=R\tilde v_\theta/\ell\gg \tilde v_\theta$, we see that $\tilde v_\theta$ and $\partial_x \tilde v_\theta$ are negligible compared to $\partial_x^2\tilde v_\theta$, which further simplifies the force balance equation. Expressing the resulting equation in terms of $r$ and $v_\theta$, we obtain
\begin{align}
\ell^2\partial_r^2 v_\theta - \frac{2\sin(2\psi_0)}{R\xi}\zeta\Delta\mu &= v_\theta,
\end{align}
where the friction length $\ell$ is given by Eq.~\eqref{eq:frictionLength}.

The solution is
\begin{align}
v_\theta&=-\frac{2\zeta\Delta\mu}{R\xi} \sin(2\psi_0)+E e^{(r-R)/\ell}  \label{eq:solvel4}
\end{align}
for $r\in(R-\ell,R)$. In this expression, we have neglected for simplicity the subdominant term proportional to $e^{-(r-R)/\ell}$. The integration constant $E$ is fixed by the boundary condition \eqref{eq:boundarycon3}. In the limit $\ell\ll R$ this condition takes the form
\begin{align}
\sigma_{\theta r}|_{r=R}&\approx \left(\eta+\frac{\gamma\tan{(2\psi_0)}^2}{4}\right) \partial_r v_\theta|_{r=R}-\frac{\zeta\Delta\mu}{2} \sin(2\psi_0) \\
\intertext{such that}
E&=\frac{2\zeta\Delta\mu  \sin(2\psi_0)\ell}{4\eta+\gamma\tan{(2\psi_0)}^2}.
\end{align}
We have verified numerically that the solution given by Eqs.~\eqref{eq:solvel3} and \eqref{eq:solvel4} is valid for $\eta\ll \gamma\tan{(2\psi_0)}^2$ and $\eta\gg \gamma\tan{(2\psi_0)}^2$, see Fig.~\ref{fig2}c,d.

\subsection{Cell number density}

To obtain the cell number density profile, we use force balance in the radial direction, Eq.~\eqref{eq:forcebalance1}. We first compute the components of the total stress tensor. The components of the Ericksen stress are given by Eqs.~\eqref{eq:StressEricksenParrr}-\eqref{eq:StressEricksenPartt}, where the terms proportional to $B$ dominate if $R^2\ll\mathcal{K}/\chi$. The antisymmetric components of the deviatory stress vanish and its symmetric components are given by Eqs.~\eqref{eq:devStressTensorRR}. 

From now on, we focus on Regimes I and II. With expression~\eqref{eq:strainratespiral} for $v_{r\theta}$ we then obtain for the total stress
\begin{align}\label{eq:devstresstensorspiral}
\sigma^\mathrm{tot}_{rr}&=\frac{B}{2}\left(1-\frac{n^2}{n_0^2}\right)-\left(\frac{1}{2}-\nu'\overline{\gamma} \frac{r^2}{R^2}\right)\frac{ \cos{(2\psi_0)}\frac{r^2}{R^2} }{1+\overline{\gamma} \frac{r^2}{R^2}}\zeta\Delta\mu - \frac{r^2}{R^2}\zeta''\Delta\mu \\
\sigma^\mathrm{tot}_{r\theta}&= \sigma^\mathrm{tot}_{\theta r}=0\\
\sigma^\mathrm{tot}_{\theta\theta}&=\frac{B}{2}\left(1-\frac{n^2}{n_0^2}\right)+\left(\frac{1}{2}+\nu'\overline{\gamma} \frac{r^2}{R^2}\right)\frac{ \cos{(2\psi_0)}\frac{r^2}{R^2} }{1+\overline{\gamma} \frac{r^2}{R^2}}\zeta\Delta\mu- \frac{r^2}{R^2}\zeta''\Delta\mu,
\end{align}
where $\overline{\gamma}=\gamma \tan{(2\psi_0)}^2/4\eta$. 

Using the above expressions in the radial component of the force balance Eq.~\eqref{eq:forcebalance1}, we can integrate once and obtain
\begin{align}
\sigma_{rr}^\mathrm{tot}&=\sigma_{rr,0}^\mathrm{tot}-\frac{r^2}{2R}\cos{(\psi_0)}T_0 +\frac{\cos(2\psi_0)}{2\overline{\gamma}}\ln{\left(\frac{1+\overline{\gamma} \frac{r^2}{R^2}}{1+\overline{\gamma}}\right)\zeta\Delta\mu}. \label{eq:totaltensionrrspirals}
\end{align}
Here $\sigma_{rr,0}^\mathrm{tot}$ is an integration constant that is fixed by the boundary condition~\eqref{eq:boundarycon4}. We now assume that the cell density deviates only little from the reference density, $|n-n_0|\ll n_0$. Equating expressions \eqref{eq:devstresstensorspiral} and \eqref{eq:totaltensionrrspirals} for $\sigma^\mathrm{tot}_{rr}$ and writing the total cell number in the circular island as $n^\mathrm{tot}\pi R^2$, we obtain up to first order in $n/n_0$
\begin{align}
\frac{n-n^\mathrm{tot}}{n_0}&\approx\frac{1}{B}\left\{\left(\frac{r^2}{R^2}-\frac{1}{2}\right)\left[\frac{R}{2}\cos{(\psi_0)}T_0 -\zeta''\Delta\mu\right]\right.\nonumber\\
&\quad\quad\left.-\frac{\cos(2\psi_0)}{2\overline{\gamma}}\left[\frac{(1-2\nu'\overline{\gamma}\frac{r^2}{R^2})\overline{\gamma} \frac{r^2}{R^2}}{1+\overline{\gamma}\frac{r^2}{R^2}}
+\ln{\left({1+\overline{\gamma} \frac{r^2}{R^2}}\right)+\Gamma}\right]\zeta\Delta\mu\right\},\label{eq:densityspirals2}
\end{align}
where $\Gamma=\nu'(\overline{\gamma}-2)-(1-2\frac{\nu'}{\overline{\gamma}})\ln{(1+\overline{\gamma})}$. Note that unlike the case of asters the density profiles of spirals depend on couplings between the field $\mathbf{h}$ and flow gradients through $\nu'$.

In the limits $\overline{\gamma}\to0$ and $\overline{\gamma}\to\infty$ we have
\begin{align}
\frac{n-n^\mathrm{tot}}{n_0}& \approx\frac{1}{B}\left(\frac{R}{2}\cos(\psi_0)T_0 -\kappa\cos(2\psi_0)\zeta\Delta\mu-\zeta''\Delta\mu\right)\left(\frac{r^2}{R^2}-\frac{1}{2}\right).
\end{align}
Here, the constant $\kappa=1$ for $\overline{\gamma}\to0$ and $\kappa=-\nu'$ for $\overline{\gamma}\to\infty$. In these limiting cases, we thus have parabolic density profiles, which differ from the cell number density for asters, Eq.~\eqref{eq:densityasters2}, only in a global pre-factor. 

\subsection{Force densities}
\label{sec:forceDesnities}

We end the discussion of spirals by determining the momentum that the monolayer exchanges with the environment in this configuration. As in the previous section, we consider only the Regimes I and II, where friction between the monolayer and the substrate is negligible. The force exerted by the monolayer on the substrate is
\begin{align}
\mathbf{t}&=-T_0\cos(\psi_0)\frac{r}{R}\mathbf{\hat{r}}.
\end{align}
At the confinement boundary $r=R$ and to first order in $n^\mathrm{tot}/n_0$, the local force density $\mathbf{f}_o$ per unit length is
\begin{align}
\mathbf{f}_o &= -\mathsf{\sigma}^\mathrm{tot}(r=R)\cdot\hat{\mathbf{r}}\\
&=\left[\frac{R}{4}\cos{(\psi_0)}T_0 
-\frac{ \nu'}{\nu} \left(\frac{\overline{\gamma}-2}{2\overline{\gamma}}+\frac{\ln{(1+\overline{\gamma})}}{\overline{\gamma}^2}\right)\zeta\Delta\mu
+\frac{\zeta''}{2}\Delta\mu
+B\left(\frac{n^\mathrm{tot}-n_0}{n_0}\right)\right]\hat{\mathbf{r}}. \label{eq:forceouterspiral}
\end{align}
As there are no azimuthal components of the force densities, the total force and torque on the system vanish, Eqs.~\eqref{eq:Ftotal} and \eqref{eq:Mtotal}.

In presence of a small friction term, the force exerted by the monolayer on the substrate now is $\mathbf{t}=-T_0\cos(\psi_0)r\mathbf{\hat{r}}/R+\xi v_\theta \hat{\bm{\theta}}$, which implies the presence of local forces and torques. The velocity $v_\theta$ is given by Eq.~\eqref{eq:solvel1} in Regime I and by Eq.~\eqref{eq:solvel2} in Regime II. The total force, Eq.~\eqref{eq:Ftotal}, still vanishes due to symmetries, whereas the total torque~\eqref{eq:Mtotal}, vanishes because the contributions from clockwise and counter-clockwise flows compensate each other.

We can generalize expression~\eqref{eq:forceinneraster} for the force exerted by the monolayer on a pillar in the center of the island obtained for asters to the case of  spirals. Making the same assumptions as in Sect.~\ref{sec:asters}, we have
\begin{align}
\mathbf{f}_i&=\mathsf{\sigma}^\mathrm{tot}(r)\cdot\hat{\mathbf{r}}\\
& =\left\{-\frac{R}{2}\cos(\psi_0)T_0 \left(\frac{r^2}{R^2}-\frac{1}{2}\right)\right.\nonumber\\
&\quad\quad\quad+\frac{\cos(2\psi_0)}{2\overline{\gamma}}\left[\ln\left(1+\overline{\gamma}\frac{r^2}{R^2}\right)+\frac{2\nu'-\overline{\gamma}}{\overline{\gamma}}\ln(1+\overline{\gamma})+\nu'(\overline{\gamma}-2)\right]\zeta\Delta\mu\nonumber\\
&\quad\quad\quad\left.-\frac{\zeta''}{2}\Delta\mu-B\left(\frac{n^\mathrm{tot}-n_0}{n_0}\right)\right\}\hat{\mathbf{r}}
\end{align}
In Regimes I and II we obtain parabolic force profiles similar to the case of asters, see Eq.~\eqref{eq:forceinneraster}, with rescaled coefficients. Note that similarly to the cell number density, the force on the pillars depends on the coupling between the field $\mathbf{h}$ and flow gradients via $\nu'$.

\section{Characterization of myoblast monolayers}

We now use the framework developed above to analyze monolayers of C2C12 myoblasts. To determine their physical properties, we analyze two different situations. First, we study the organization of cells around topological defects in extended confluent layers. Through our analysis, we constrain the Frank elastic constants, which characterize splay and bend deformations of the orientational order field. Second, we examine spiral arrangements of monolayers confined to small circular domains. This analysis allows us to comprehensively determine the material parameters of myoblast monolayers. For experimental details, we refer to Ref.~\cite{PauScience}.

\subsection{Nematic elastic moduli}\label{sec:FrankConstants}

In the following we determine the ratio of the nematic elastic constants for extended confluent C2C12 monolayers. In this situation, the cells exhibit long-ranged orientational order and arrange into patterns similar to passive nematic liquid crystals~\cite{Duclos2017}. The nematic organization is evidenced for instance by the presence of half-integer topological defects~\cite{PauScience}. We capture the nematic order by  the director field $\mathbf{n}$ and analyze its configurations around +1/2 topological defects in terms of an equilibrium approach to nematic liquid crystals. Similar approaches were used in the context of synthetic or biological liquid crystals~\cite{Brugues2008,Zhang2017}. 

For a two-dimensional nematic liquid crystal with director field $\mathbf{n}$, the elastic energy associated with distortions of the orientational order is
\begin{align}
\mathcal{F}&=\int_\mathcal{A}\left\{\frac{\mathcal{K}_1}{2}\left(\nabla\cdot \mathbf{n}\right)^2+\frac{\mathcal{K}_3}{2}\left(\mathbf{n}\times\left(\nabla\times\mathbf{n}\right)\right)^2\right\}da \label{eq:freeenergyanisotropic}
\end{align}
with Frank elastic constants $\mathcal{K}_1$ and $\mathcal{K}_3$. They, respectively, quantify the energetic costs of splay and bend deformations~\cite{DeGennes1995}.

\begin{figure}[b] %  figure placement: here, top, bottom, or page
	\centering
	\includegraphics[]{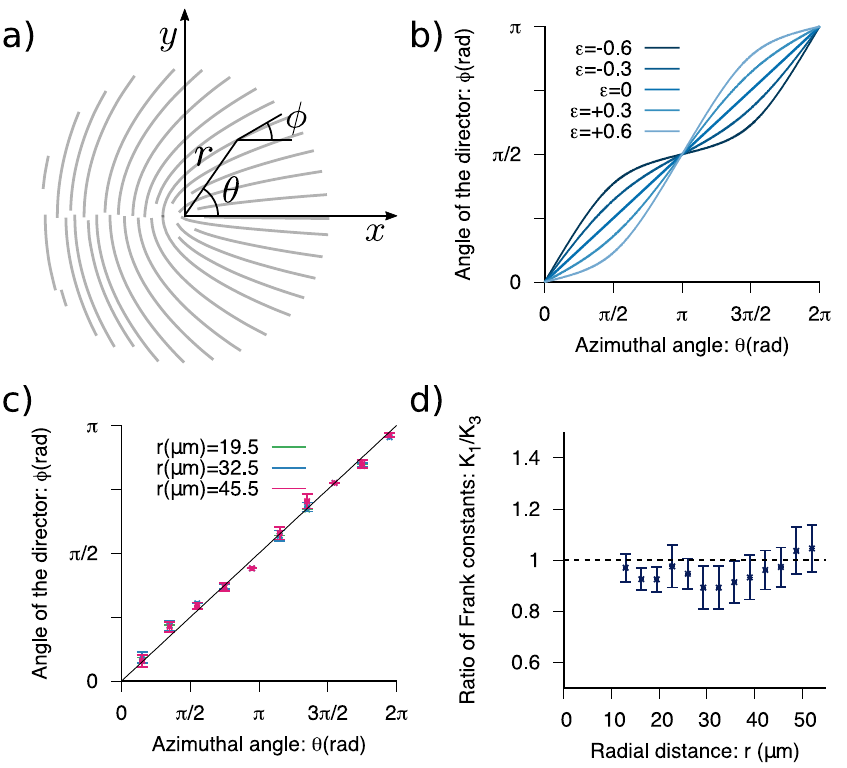}
	\caption{(online color) Half-integer topological defects in C2C12 myoblast monolayers. a) Schematic representation of the director field for a $+1/2$ topological defect. b) Theoretical profile $\phi(\theta)$, Eq.~\eqref{eq:directorfield}, with $s=+1/2$ for varying $\epsilon$ as indicated in the legend. The ratio of Frank constants is: $\mathcal{K}_1/\mathcal{K}_3=\{0.25,0.54,1,1.86,4.\}$ for $\epsilon=\{-0.6,-0.3,0,0.3,0.6\}$. c) Representative experimental curves $\phi^e(\theta)$ for varying radial distance $r$ as indicated in the legend. d) Fitted ratio $\mathcal{K}_1/\mathcal{K}_3$ as a function of the radial coordinate $r$. Error bars correspond to the std of all values of $\epsilon$ that lead to $\mathcal{E}<1.1\mathcal{E}_{min}$.}  \label{figS1}
\end{figure}

The equilibrium director configuration is determined by minimizing the energy \eqref{eq:freeenergyanisotropic}. Near a topological defect, the solution is given by~\cite{dzyaloshinskii1970theory}
\begin{align}
\theta&=p\int_0^{\phi-\theta}\sqrt{\frac{1+\epsilon \cos{(2x)}}{1+\epsilon p^2 \cos{(2x)}}}dx,  \label{eq:directorfield}
\end{align}
where the elastic anisotropy parameter is $\epsilon=(\mathcal{K}_1-\mathcal{K}_3)/(\mathcal{K}_1+\mathcal{K}_3)$, for which there is a one-to-one correspondance with the ratio ${\cal K}_1/{\cal K}_3$. Furthermore, $\phi$ denotes the angle of the director $\mathbf{n}$ with respect to a fixed axis and $\theta$ is the azimuthal angle with respect to the defect center, Fig.~\ref{figS1}a. The fixed axis is chosen such that $\phi(\theta=0)=0$. Note that Eq.~\eqref{eq:directorfield} is independent of the radial coordinate $r$, Fig.~\ref{figS1}a. Finally, $p$ is a constant that is determined by the condition that $\phi$ is a single-valued function of $\theta$, which leads to
\begin{align}
\pi=(s-1)p\int_0^\pi\sqrt{\frac{1+\epsilon \cos{(2x)}}{1+\epsilon p^2 \cos{(2x)}}}dx, \label{eq:boundarycondition}
\end{align}
where $s$ corresponds to the topological charge of the defect. Figure~\ref{figS1}b shows $\phi(\theta)$ for a $s=+1/2$ topological defect and for varying $\epsilon$.

For extended C2C12 monolayers, we obtained the experimental values $\phi^e$ by first determining the director field of the monolayer using structure factor methods~\cite{Puspoki2016}, see Methods in Ref.~\cite{PauScience}. We then averaged the director orientation over time for $N>100$ distinct $+1/2$ topological defects. For the overall average, we fixed the radial coordinate $r$ and thus obtained average profiles for different radial distances, see Fig~\ref{figS1}c. Within the experimental error, the director orientation did not depend on $r$, which is in agreement with the theory. We fitted the solution \eqref{eq:directorfield} for $\phi$ to the experimental data by using the elastic anisotropy $\epsilon$ as the only fit parameter. The parameter $\epsilon$ was obtained by minimizing the error function
\begin{align}
\mathcal{E}=\int_0^{2\pi}|\phi(\theta)-\phi^e(r,\theta)|d\theta.\label{eq:funcioerror}
\end{align}
We attributed an error to this value as the standard deviation (std) of all values of $\epsilon$ that lead to $\mathcal{E}<1.1\mathcal{E}_{min}$, where $\mathcal{E}_{min}$ is the absolute minimum. 

The values of $\mathcal{K}_1/\mathcal{K}_3$ thus obtained are presented in Fig.~\ref{figS1}d as a function of the radial distance $r$ with respect to the defect center. Although there is some tendency of the ratio $\mathcal{K}_1/\mathcal{K}_3$ to increase with $r$, there is not a significant difference between the values of this ratio for different radii. The value averaged over all experimental data is $\mathcal{K}_1/\mathcal{K}_3=0.95\pm 0.10$ (mean$\pm$std). We conclude that the Frank elastic constants $\mathcal{K}_1$ and $\mathcal{K}_3$ are equal within the experimental error. This justifies our choice of the one-constant approximation made in Eq.~\eqref{eq:freeenergy}, where $\mathcal{K}=\mathcal{K}_1=\mathcal{K}_3$.

\subsection{Determination of material parameters}\label{sec:Fittingprocedure}

In order to determine the material parameters of C2C12 myoblast monolayers, we solve the full dynamic equations for a broad range of parameters numerically, see App.~\ref{sec:numerics}, and compare the velocity and polarization fields obtained in this way to our experimental data. Specifically, we used data from spirals on islands with radius $R=50~\mu$m, 100~$\mu$m, and 150~$\mu$m for the velocity $v_\theta$ and the polar order parameter $S$. For the polarization angle $\psi$, we used data from spirals on islands with a fixed radius $R=100~\mu$m. 

The difference between the numerical and experimental fields are quantified via an error function $\mathcal{E}$ that are given below. The parameter set that gives the minimal error $\mathcal{E}_\mathrm{min}$ then provides the sought for material parameters. We will determine confidence intervals for these parameter values by considering the range of parameter values that yield an error within 10\% of the minimal error, that is, for which $\mathcal{E}<1.1\mathcal{E}_\mathrm{min}$. 

\begin{figure}[t] %  figure placement: here, top, bottom, or page
\centering
\includegraphics[]{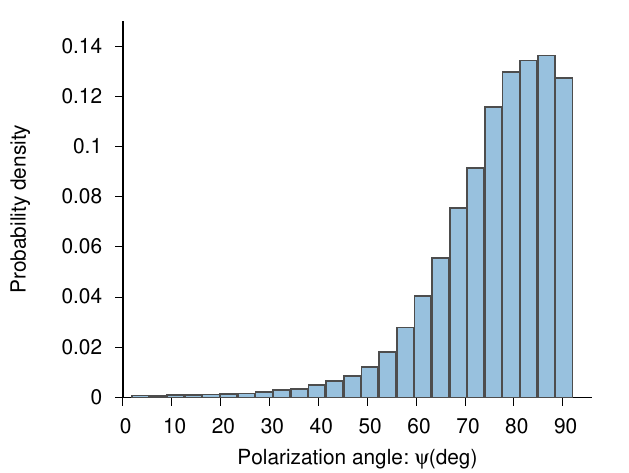}
\caption{(online color) Probability density of the polarization angle with respect to the radial direction $\psi$. The data was obtained from C2C12 monolayers in spiral configurations that were confined to an island of $100$~$\mu$m radius ($N=12$).}\label{fig3a}
 \end{figure}

The numerical solutions are computed after making the dynamic equations dimensionless. To this end, we use the radius $R$ of the smallest island as the length scale, $\mathcal{K}$ as the energy scale, and $\mathcal{K}/(R\gamma)$ as the velocity scale. The flow alignement parameter $\nu=1/\cos(2\psi_0)$ can be directly inferred from the angle $\psi=\psi_0$ between the polarization vector and the radial direction, Fig.~\ref{fig3a}. The average angle $\psi=76\pm13^\circ$, which leads to $\nu=-1.1\pm0.3$ (mean$\pm$std, $N=12$). For the numerical calculations, we used $\nu=-1.2$. This leaves us with 5 dimensionless parameters to determine: $\chi R^2/\mathcal{K}$, $\eta/\gamma$, $\xi R^2/\gamma$, $\zeta\Delta\mu R^2/\mathcal{K}$, and $T_0R^3/\mathcal{K}$. In the remainder of this section, we will use the same notation for the nondimensionalized parameters as for the original ones.

\begin{figure}[t] %  figure placement: here, top, bottom, or page
\centering
\includegraphics[]{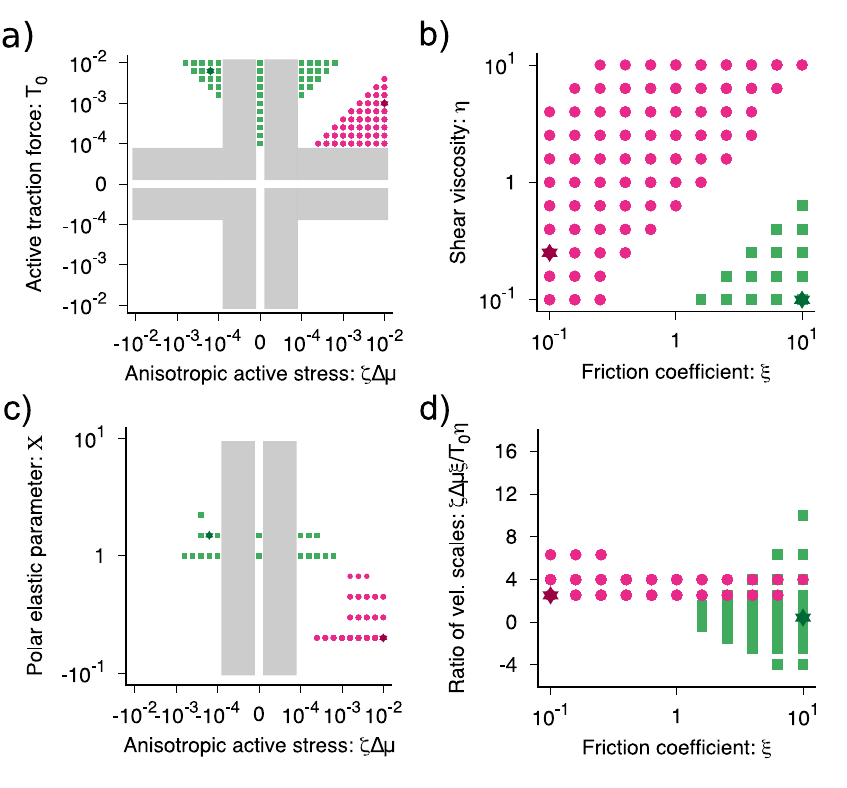}
\caption{(online color) Parameter values leading to an error $\mathcal{E}<1.1\mathcal{E}^{min}$ for the error function~\eqref{eq:errorfunction}. The cuts of the parameter space are: a) $T_0$ vs $\zeta\Delta\mu$, b) $\eta$ vs $\xi$, c) $\chi$ vs $\zeta\Delta\mu$, and d) $\zeta\Delta\mu\xi/T_0\eta$ vs $\xi$. The units are fixed by $\mathcal{K}=\gamma=R=1$, and $\nu=-1.2$. Gray areas indicate parameter regions that were not analyzed. Green squares: active stress dominated region, dark green star: local minimum. Magenta circles: traction force dominated region, dark magenta star: global minimum. }\label{fig4}
 \end{figure}

We computed solutions for parameters in the range $(\chi, \eta,\xi,|\zeta\Delta\mu|,|T_0|)\in(0.2,5)\times(10^{-1},10^1)\times(10^{-1},10^1)\times(10^{-4},10^{-2})\times(10^{-4},10^{-2})$, where $\zeta\Delta\mu$ and $T_0$ can take either sign. As error function we used 
\begin{align}
\mathcal{E}&=\sum_i |v_{\theta,i}^{e}-v_{\theta,i}|\Delta r_i+\sum_i|S_i^{e}-S_i| \Delta r_i.\label{eq:errorfunction}
\end{align}
Here, the superscript 'e' indicates values averaged over at least $N=5$ experiments, and the index $i$ indicates that samples are taken at discrete radial positions $r_i$. Furthermore, $\Delta r_i=r_{i+1}-r_i$ is related to the experimental spatial resolution and $\Delta r_i\sim 5~\mu$m. In Figure~\ref{fig4}, we present various cuts through the parameter space and indicate the regions, where $\mathcal{E}<1.1\mathcal{E}_\mathrm{min}$.
 
 \subsection{Myoblast monolayers confined to circular domains}
 \label{sec:monolayers}
 
In this section, we discuss the parameter values determined by the approach described in the previous section using our experiments of C2C12 monolayers on circular domains~\cite{PauScience}. Let us start by setting the units of our experiments. The length scale is set by the radius of the smallest island $R=50~\mu$m. The velocity scale is set by the azimuthal flow velocity at the edges of the island to $30~\mu$m/h. Finally, the energy scale is set by the stress exerted on pillars of radius 40~$\mu$m times $R^3$, that is, 10~kPa$\times1.25\cdot10^5~\mu$m$^3=1.25\cdot10^3~\mu$N$\mu$m.

The data presented in Figure~\ref{fig4} readily reveals several constraints on the parameter values. First of all $T_0>0$, Fig.~\ref{fig4}a, which shows that the azimuthal velocity $v_\theta$ is in the direction of the azimuthal component of the polarization field $\mathbf{p}$. Second, the penetration length of the polar order parameter $\sqrt{{\cal K}/\chi}$ is larger than $25$~$\mu$m, Fig.~\ref{fig4}c. It is thus at least of the same order as the confinement radii in our experiments, such that the orientational order induce by the boundaries propagates into the center of the island.

\begin{figure}[t] %  figure placement: here, top, bottom, or page
\centering
\includegraphics[]{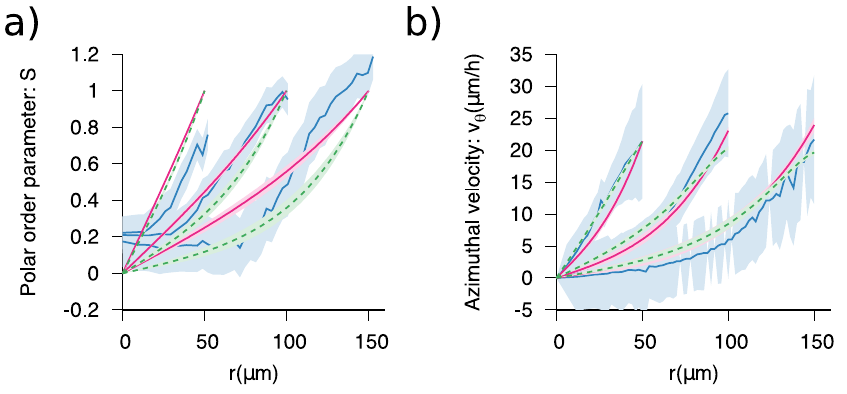}
\caption{(online color) Theoretical fits to experimental data. a) Polar order parameters $S$ and b) azimuthal velocity $v_\theta$ as a function of the radial distance $r$. Mean theoretical profiles for the active stress dominated parameter region in solid magenta and for the traction force dominated parameter region in dashed green, see Fig.~\ref{fig4} and Table~\ref{tab:table1}. Blue: experimental profiles ($N=(11,12,5)$ for confining domain radius $(50,100,150)$~$\mu m$). Error bars in theoretical fits correspond to the std of parameter values that lead to $\mathcal{E}<1.1\mathcal{E}_{min}$ and in experimental curves to sem. Profiles for three different confinement radii $R=50$, $100$, and $150$~$\mu$m are shown. The theoretical curves are endowed with physical units such that $S(R)=1$ and $v_\theta(R)=21.4~\mu$m/h for $R=50~\mu$m.}\label{fig5}
\end{figure}

Further inspection of Fig.~\ref{fig4} shows two disjoint region in parameter space corresponding to solutions with distinct physical properties. In both cases, the parameters yield close fits to the polar order parameter $S$ and the azimuthal velocity $v_\theta$ measured in our experiments, see Fig.~\ref{fig5}. The two regions are narrow in several directions, meaning that the corresponding combinations of the dimensionless parameters are well determined by our experimental data. This is the case, for example, for $\zeta\xi R/T_0 \eta$, see Fig.~\ref{fig4}b and Table~\ref{tab:table1}. The directions that are less constrained still provide upper or lower bounds on our dimensionless parameters, see Table~\ref{tab:table1}.
\begin{table}[b]%The best place to locate the table environment is directly after its first reference in text
\begin{ruledtabular}
\begin{tabular}{lcc}
 &Active stress& Traction force \\
 &  dominated & dominated\\
$T_0 R/|\zeta\Delta\mu|$ with $T_0>0$ & $<0.6^*$ &  $>16$ \\
$\sqrt{\eta/\xi}/R$ & $>0.5$ &$<0.24$  \\
$\sqrt{{\cal K}/\chi}/R$ & $>1$ & $(0.4,2)$ \\
$\zeta\Delta\mu R\xi/\eta T_0 $ with $T_0>0$ & $3.2\pm 1.3$ & $0.5\pm1.4$ \\
$\nu$ & $-1.1\pm0.3 $ & $-1.1\pm0.3 $ 			
\end{tabular}
\end{ruledtabular}
\caption{Table of material parameters for the solutions in Fig.~\ref{fig5}. The errors correspond to std. To restore length units $R=50$~$\mu$m. $^*$ with $\zeta>0$. \label{tab:table1}}	
\end{table}

The parameter region for the solid magenta fits in Fig.~\ref{fig5} corresponds to a mechanical regime where the anisotropic active stress $\zeta\Delta\mu$ is the dominating active mechanism, $T_0R/|\zeta\Delta\mu|< 0.6$. In this active stress dominated regime, the length scale $\sqrt{\eta/\xi}$, which is determined by the dissipative mechanisms, is bounded from below by 25~$\mu$m. The penetration length of the polar order is $\sqrt{\mathcal{K}/\chi}>50$~$\mu$m. There are two velocity scales associated with the two active mechanisms, $\zeta\Delta\mu R/\eta$ and $T_0/\xi$. The ratio between these two scales $\zeta\Delta\mu R\xi/\eta T_0 =3.2\pm 1.3$ shows that the flows are mainly generated by anisotropic active stresses.

The parameter region for the dashed green fits in Fig.~\ref{fig5} corresponds to a mechanical regime, where the traction force $T_0$ is the dominating active mechanism, $T_0R/|\zeta\Delta\mu|> 16$. In this traction force dominated regime, the length scale $\sqrt{\eta/\xi}$ is bounded from above by $12$~$\mu$m. The penetration length of the polar order is limited $22~\mu$m$<\sqrt{\mathcal{K}/\chi}<112$~$\mu$m. The ratio of the two velocity scales $\zeta\Delta\mu R\xi/\eta T_0 =0.5\pm 1.4$ shows that the flows are mainly generated by traction forces.

\begin{figure}[t] %  figure placement: here, top, bottom, or page
\centering
\includegraphics{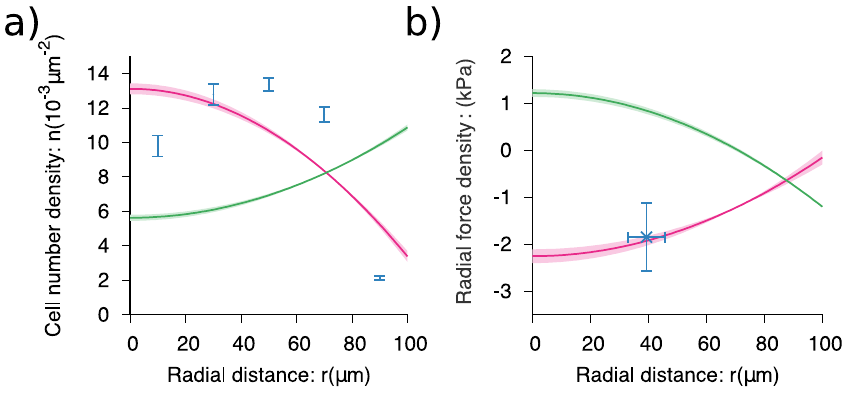}
\caption{(color online) Theoretical fits of steady state profiles for asters. a) Cell number density $n$, b) radial force density as a function of the radial distance $r$. Averaged experimental profiles (blue, $N=10$ in (a) and $N=3$ in (b)), mean fit in the active-stress dominated (magenta, full lines) and in the traction dominated parameter region (green, dashed lines). The theoretical solutions are Eq.~\eqref{eq:densityasters2} in (a) and Eq.~\eqref{eq:forceinneraster} in (b). Parameters are given in Tab.~\ref{tab:table2}. We used $\zeta''\Delta\mu=0$. Error bars in theoretical fits correspond to std of all parameter values with $\mathcal{E}<1.1\mathcal{E}_{min}$ and in experimental curves to sem. In Fig.~3 of Ref.~\cite{PauScience}, the compressional stresses correspond to minus the radial force density in panel (b).}\label{fig6} 
\end{figure}

Although, the two parameter regions give comparably good fits to the polar order parameter and the azimuthal velocity in spirals, their mechanical characteristics are distinct. An important difference between the two regions is exhibited in the steady state force density and cell number density profiles of asters. In the active stress dominated region, the cell number density increases towards the center whereas it decreases towards the center in the traction force dominated regime, see Fig.~\ref{fig6}a. 

Furthermore, the force density is pointing towards the center of the circular domain in the active stress dominated region, whereas it is pointing outwards in the traction force dominated region, see Fig.~\ref{fig6}b. In our experiments, we observed an increase of the cell number density in the center compared to the periphery, see Fig.~3 in Ref.~\cite{PauScience}. A further sign of cell accumulation in the center was the formation of mounds, see Figs.~1, 4 in Ref.~\cite{PauScience}. When elastic pillars were placed in the center of the circular domain, we observed compression of these structures, which is again compatible with the active stress dominated region, see Fig.~3 in Ref.~\cite{PauScience}. 

\begin{table*}[t]
\begin{ruledtabular}
\begin{tabular}{cccccccc}
$T_0(\text{Pa})$ & $\zeta\Delta\mu(\text{kPa}~\mu\text{m})$ & $\eta(\text{kPa h}~\mu\text{m})$ & $\xi(\text{Pa h}/\mu\text{m})$ & $\sqrt{{\cal K}/\chi}(\mu\text{m})$ & $\nu$ & $B/n_0(\text{kPa}~\mu\text{m}^3)$ & $n^{\mathrm{tot}}(10^{-3}~\mu\text{m}^{-2})$ \\
 $<600\pm60$ & $48\pm4$ & $34\pm8$ & $<40\pm20$ & $>50$ & $-1.1\pm0.3$ & $4600\pm800$ & $8.2\pm0.5$ 
  			
\end{tabular}
\end{ruledtabular}
\caption{Table of material parameters for active stress dominated solutions. To convert 3d material parameters into 2d material parameters we use a cell monolayer height of $10$~$\mu$m.  Error bars correspond to std of all parameter value with $\mathcal{E}<1.1\mathcal{E}_{min}$ except for $\nu$ (mean$\pm$std). \label{tab:table2}}	
\end{table*}

For the fits presented in Fig.~\ref{fig6}b, we have imposed that the isotropic stress $\zeta''\Delta\mu$ vanishes. If this value were used as a fitting parameter, a qualitative agreement between the theory and the experiment could be achieved in the traction-force dominated regime, such that a discrimination between the two regimes might appear not to be possible based on these fits. However, in that case, the isotropic stress $\zeta''\Delta\mu$ needs to be comparable to $T_0R$ to achieve the same order of magnitude for the stress exerted on the pillars, see Eq.~\eqref{eq:forceinneraster}. We conclude that traction forces cannot be the dominating mechanism for generating pillar deformations.

To obtained the material parameters in the active stress dominated region, Table~\ref{tab:table2}, we combined the analysis from the polarization and velocity fields in spirals, Fig.~\ref{fig5}, with the cell number density and stresses fields in asters, Fig.~\ref{fig6}. Specifically, we restored the velocity units by setting $v_\theta(r=R)=21.3$~$\mu$m/h for $R=50$~$\mu$m and obtained the ratio $\zeta\Delta\mu /\eta=1.4\pm0.3$~h$^{-1}$. With a similar fitting procedure to that explained in Sec.~\ref{sec:Fittingprocedure}, we fitted the theoretical steady state profiles for asters, Fig.~\ref{fig6}, and obtained the parameters $B/n_0$, $n^\mathrm{tot}$, and $\zeta\Delta\mu$ listed in Table~\ref{tab:table2}. To transform the stress that cells exerted on deformable pillars into 2d cell monolayer stresses, we considered that the height of the monolayer was $10$~$\mu$m. Combining these new results with those from Table~\ref{tab:table1}, we obtained the material parameters from Table~\ref{tab:table2}.

\subsection{Comparison to other cell monolayers and conditions}
 
Next, we discuss how our estimates of the material parameters compare to other cellular systems or conditions. First, for contractile epithelial monolayers, $\zeta\Delta\mu<0$, an analog of a de-wetting transition was found~\cite{Perez-Gonzalez2019}. This transition was controlled by the length scale $-\zeta\Delta\mu/T_0$. In our case, such a transition is not expected to occur, because in both parameter regions the system is either dominated by  traction forces or by extensile active stresses, Table~\ref{tab:table1}.

Previous experiments had identified C2C12 monolayers as being contractile ($\zeta\Delta\mu<0$). This conclusion was drawn from the dynamics of $+1/2$ topological defects~\cite{Kawaguchi2017}. In other experiments, based on the direction of the cellular shear flows with respect to the orientation of the cell bodies, it was concluded that these monolayers are extensile ($\zeta\Delta\mu>0$)~\cite{Duclos2018}. In our experiments, the observed flows in spirals are compatible with extensile active stresses in the active stress dominated regime. In the traction force dominated regime both, contractile and extensile active stresses, were compatible with the flows, see Fig.~\ref{fig4}a. Further work is necessary to understand the difference between these experiments.

The flow-alignment parameter $\nu=-1.1\pm 0.3$ controls the re-orientation of the polarization field $\mathbf{p}$ in response to shear flows. This value is similar to the typical range for passive liquid crystals~\cite{DeGennes1995}. In the drosophila wing, this parameter was estimated to be $-1<\nu<-10$~\cite{Aigouy2010}. 

The mechanics of individual C2C12 cells was assessed by confining them to micropatterns of varying geometries~\cite{Bruyere2019}. There, it was found that traction forces of elongated C2C12 cells were concentrated at the distal ends of the cell body and pointed inwards. Depending on the cell geometry, these corresponding stresses ranged between $100$ and $1000$~Pa. For monolayers of other elongated cell types, the force per unit length associated with intracellular interactions were of the order of 10 kPa $\mu$m~\cite{Vincent2015}. In our experiments, we observed that confluent monolayers compressed elastic pillars with a stresses of the order of 1-10~kPa.

For spreading epithelial monolayers, the friction length was found between $100$ and $1000$~$\mu$m~\cite{Blanch-Mercader2017,Moitrier2019}. Such large values result from stable cell-cell junctions formed by epithelial cells. For cell types lacking such junctions, like C2C12 myoblasts, the friction length was found to be smaller, $10-40$~$\mu$m~\cite{Duclos2018}. The latter values are of the same order of magnitude as the bounds we found in both parameter regions for $\sqrt{\eta/\xi}$, which is smaller than the friction length $\ell$ given by Eq.~\eqref{eq:frictionLength}, see Table~\ref{tab:table1}.

Also the penetration length of the polarity field $\sqrt{{\cal K}/\chi}$ was measured in epithelial monolayers~\cite{Blanch-Mercader2017,Perez-Gonzalez2019}. It was found to be between $10$ and $100$~$\mu$m, which is of the same order as in our measurements. When epithelial monolayers were confined to circular islands with radii comparable to $\sqrt{{\cal K}/\chi}$, collective rotation was found~\cite{Doxzen2013,Deforet2014,Segerer2015}. However, in these cases, no evidence of topological defects organizing these flows was reported.

\section{Extensions}

In this section, we discuss the effects of extensions to our dynamical system. In particular, we consider nematic traction forces and active alignment.

\subsection{Nematic traction forces}\label{sec:ActiveNematicForce}

In the force balance Eq.~\eqref{eq:forcebalance}, we considered the active forces exerted by the monolayer onto the substrate result from processes with polar symmetry, $T_0\mathbf{p}$. In principle, also processes with nematic symmetry, which remain invariant under the operation $\mathbf{p}\rightarrow-\mathbf{p}$, could contribute to these forces. In some cases, these contributions have been shown to be of the same order as the polar contributions~\cite{Maitra2018}. We now discuss the effects of such terms on spirals and asters.

Up to second order in $\mathbf{p}$ and first order in derivatives, the nematic contributions to the right hand side of the force balance equation~\eqref{eq:forcebalance} can be written as
\begin{align}
\partial_\beta \left(p_\alpha p_\beta-\frac{1}{2}p_\gamma p_\gamma \delta_{\alpha\beta}\right)T_1+ \partial_\beta\left(p_\gamma p_\gamma \delta_{\alpha\beta}\right)T_2+\left( p_\alpha \partial_\beta p_\beta- p_\beta \partial_\beta p_\alpha\right)T_3. \label{eq:App2}
\end{align}
Addition of the first two terms to the force balance equation amounts to a redefinition of the coupling coefficients $\zeta$ and $\zeta''$ in the constitutive equation~\eqref{eq:devstresstensor} for the deviatory stress, $\zeta\Delta\mu\rightarrow \zeta\Delta\mu+T_1$ and $\zeta''\Delta\mu\rightarrow\zeta''\Delta\mu+T_2$. Due to substrate interactions, a contractile system can thus become extensile or vice versa, but the terms proportional to $T_1$ and $T_2$ do not introduce qualitatively new behavior.

The antisymmetric term proportional to $T_3$, in contrast, cannot be absorbed in the constitutive equation~\eqref{eq:devstresstensor}. In principle, this term can thus lead to new effects compared to our original system. Let us evaluate its effects on spirals and asters in small confinements with $R^2\ll\mathcal{K}/\chi$. Expressing the components of $\mathbf{p}$ in terms of the nematic order $S$ and the angle $\psi$ of the director with the radial direction, it reads
\begin{align}
\left(\frac{S^2}{r}\hat{\mathbf{r}}-S^2\partial_r\psi \hat{\bm{\theta}}\right)T_3 \label{eq:App3}.
\end{align}
For the steady-state spirals and asters considered above, we have $S=r/R$ and $\psi=const$, such that the term reduces to $T_3 r\hat{\mathbf{r}}/R^2$, which has the same form as the term proportional to $\zeta''\Delta\mu$ on the left hand side of the force balance equation~\eqref{eq:forcebalance1}. We conclude that nematic traction forces do not introduce new effects in spirals and asters aside from possibly introducing additional surface terms. 

\subsection{Active alignment}\label{sec:activeAlignment}

In the constitutive equation for the dynamics of the polarization field, Eq.~\eqref{eq:dinamicadirector}, we have neglected a coupling to the chemical thermodynamic force $\Delta\mu$. Explicitly, the term would be of the form $\mathbf{p}\lambda\Delta\mu$. Depending on the sign of the phenomenological constant $\lambda$, this term favors the generation or inhibition of polar order by active processes~\cite{Julicher2007}. Note that this 'active alignement' is different from spontaneously emergent orientational order by active flows~\cite{Mueller2019,Santhosh2020}.

For our choice of the free energy, see Eq.~\eqref{eq:freeenergy}, the molecular field $\mathbf{h}$ contains a term $-\chi\mathbf{p}$, such that in the dynamic equation~\eqref{eq:dinamicadirector}, the presence of active alignment can be absorbed into the parameter $\chi$ such that $\chi\rightarrow\chi-\gamma\lambda\Delta\mu$.  Due to activity, the sign of the redefined $\chi$ can thus be different from that of $\chi$. However, because C2C12 monolayers confined to small circular domains exhibit a disorganized center, the pre-factor of $\mathbf{p}$ in Eq.~\eqref{eq:dinamicadirector} should be positive, as in our above analysis.

A redefinition of the parameter $\chi$ also affects the symmetric part of the deviatory stress tensor, Eq.~\eqref{eq:devstresstensor} and the Ericksen stress tensor, Eq.~\eqref{eq:Ericksencomplete}. These effects can be absorbed by a redefinition of the coupling coefficients $\zeta$ and $\zeta''$. Explicitly, $\zeta\rightarrow\zeta+\nu\lambda\gamma$, and $\zeta''\rightarrow\zeta''+\lambda\gamma(\nu'-1/2)$. We conclude that an active alignment term in the dynamic equation for the polarization field $\mathbf{p}$ does not qualitatively change the behavior of our system aside from possibly introducing additional surface terms.
 
\section{Discussion}

In summary, we have analyzed in detail the steady state patterns of spirals and asters of a compressible active polar fluid. We showed that isolated topological defects provide information for quantifying material parameters of cell monolayers. Small circular confinements allowed us to control the position and topological charge of such defects. In principle, other techniques could be used for this purpose, in particular, micropatterning of the topography of the substrate~\cite{Endresen2019,Turiv2020} or application of external magnetic fields~\cite{Dua1996}. These methods allow to impose spatiotemporal cell orientation patterns, which in our system were self-organized. Combining these approaches opens a vast range of possibilities to improve our quantitative understanding of cell monolayer mechanics. 

Ideally, asters and spirals in two-dimensional nematic phases exhibit a single point, where the orientational order is ill-defined. In our experiments, cell monolayers were disorganized in a central region, see Fig.~\ref{fig001}, that increased in size with the radius of the confining domain. Order was found in a region close to the domain boundary. An alternative interpretation of the steady state aster and spiral patterns considers the ordered region to be a boundary layer. Still, the same dynamic equations could be used to analyze the data, such that our results are independent of the interpretation. 

The lack of spontaneously emerging orientational order in the center of the confining domain led us to consider $\chi>0$ in the free energy \eqref{eq:freeenergy}. In extended C2C12 monolayers, however, long-range orientational order can be observed for similar cell number densities~\cite{Duclos2017,Kawaguchi2017,PauScience}. This observation suggests that in the range of domain sizes used in this work, the boundary-induced order overcomes the density-induced order. To explicitly study this competition, a description  of mixed orientation, nematic and polar, would be needed.  

Furthermore, in our experiments, asters appeared as the cell number increases, suggesting that cell number density is a control parameter for the transition. Indeed, when proliferation was inhibited in spiral configurations~\cite{PauScience}, asters were not observed. This effect is not captured by our theory and would require a better understanding of the physics underlying cell orientation at interfaces. 

Topological defects have been suggested to be involved in morphogenetic processes~\cite{Maroudas-Sacks2020}. In a similar way to our work, one could use these defects to quantify the material properties of the tissue. Such an analysis could reveal the physical conditions underlying collective cell migration during morphogenesis and provide essential pieces of information for understanding developmental processes.

\begin{acknowledgments}
We thank Zena Hadjivasiliou for suggesting the systematic parameter sampling performed in Sect.~\ref{sec:Fittingprocedure} and Jean-Fran\c cois Joanny for discussions.
\end{acknowledgments}
 
\appendix

\section{The Ericksen stress tensor}\label{sec:EricksenStress}

In the following, we compute the Ericksen stress tensor $\sigma^e_{\alpha\beta}$ for a compressible active polar fluid with the free energy \eqref{eq:freeenergy} and give the corresponding Gibbs-Duhem relation~\cite{Ericksen1962}. For a one-component polar fluid with cell number density $n$ and polarization field $p_{\alpha}$, the general expression for the Ericksen stress tensor takes the form~\cite{Joanny2007,Furthauer2012}
\begin{align}
\sigma^{e}_{\alpha\beta}=(f-n\mu)\delta_{\alpha\beta}-\frac{\partial f}{\partial (\partial_{\beta}p_{\gamma } )}\partial_{\alpha}p_{\gamma}\label{eq:Ericksen}.
\end{align}
Here, $f$ is the free energy density, such that $\mathcal{F}=\int fda$, and $\mu=\frac{\partial f}{\partial n}$ the chemical potential. With the free energy \eqref{eq:freeenergy}, we obtain
\begin{equation}
\sigma^{e}_{\alpha\beta}=\left[\frac{B}{2}\left(1-\frac{n^2}{n_0^2}\right)+\frac{\chi}{2}p_{\gamma}^2+\frac{{\cal K}}{2}(\partial_{\gamma}p_{\delta})^2\right]\delta_{\alpha\beta}-{\cal K}(\partial_{\alpha}p_{\gamma})(\partial_{\beta}p_{\gamma}).  \label{eq:Ericksencomplete}
\end{equation}

Writing the radial and the azimuthal components of the polarization vector again as $p_r=S\cos(\psi)$ and $p_\theta=S\sin(\psi)$, respectively, the components of the Ericksen stress in polar coordinates are 
\begin{align}
\sigma^{e}_{rr}&=\frac{B}{2}\left(1-\frac{n^2}{n_0^2}\right)+\frac{\chi}{2}S^2+\frac{{\cal K}}{2}\left[\frac{S^2}{r^2}-(\partial_r S)^2-S^2 (\partial_r \psi)^2 \right]\label{eq:StressEricksenParrr}\\
\sigma^{e}_{r\theta}=\sigma^{e}_{\theta r}&=-{\cal K}S^2\frac{\partial_r \psi}{r}\label{eq:StressEricksenParrt}\\
\sigma^{e}_{\theta\theta}&=\frac{B}{2}\left(1-\frac{n^2}{n_0^2}\right)+\frac{\chi}{2}S^2-\frac{{\cal K}}{2}\left[\frac{S^2}{r^2}-(\partial_r S)^2-S^2 (\partial_r \psi)^2\right],\label{eq:StressEricksenPartt}
\end{align}
where we have assumed rotational invariance.

The Gibbs-Duhem relation links the intensive variables of the free energy and reads~\cite{Joanny2007,Furthauer2012}
\begin{align}
\partial_{\gamma}\sigma^e_{\alpha\gamma}&=-n\partial_{\alpha}\mu-h_{\gamma}\partial_{\alpha}p_{\gamma}.\label{eq:GibbsDuhem}
\end{align}
In polar coordinates this expression yields
\begin{align}
\partial_{r}\sigma^e_{rr}+\frac{\sigma^e_{rr}-\sigma^e_{\theta\theta}}{r}&=-n\partial_r\mu-h_r \partial_r p_r-h_{\theta}\partial_r p_\theta\nonumber\\
&=-n \partial_r \mu-h_\parallel \partial_r S-h_\perp S \partial_r \psi\label{eq:FBStressEricksenParA}\\
\partial_{r}\sigma^e_{\theta r}+\frac{\sigma^e_{r\theta}+\sigma^e_{\theta r}}{r}&=-h_r (\frac{-p_\theta}{r})-h_\theta (\frac{p_r}{r})\nonumber\\
&=-\frac{h_\perp S}{r}\label{eq:FBStressEricksenParB},
\end{align}
where in the second step we have expressed the polarization vector in terms of $S$ and $\psi$ and used the components $h_\parallel$ and $h_\perp$ of the molecular field. These relations can be verified explicitly by inserting the expressions~\eqref{eq:StressEricksenParrr}-\eqref{eq:StressEricksenPartt} for the components of the Ericksen stress and using Eqs.~\eqref{eq:molecularfieldpara} and \eqref{eq:molecularfieldperp} for $h_\parallel$ and $h_\perp$. 

\section{The molecular field}\label{sec:molecularfield}

In the following, we compute the expression of the molecular field $\mathbf{h}$ in polar coordinates. Assuming rotational invariance of our system, the physical fields are independent of the azimuthal angle $\theta$, and the free energy \eqref{eq:freeenergy} can be written as 
\begin{align}
\mathcal{F}&=\int_{\cal A}\left\{\frac{B}{2}\left(\frac{n}{n_0}-1\right)^2+\frac{\chi}{2}(p_r^2+p_\theta^2)+\frac{{\cal K}}{2}\left[(\partial_r p_r)^2+(\partial_r p_\theta)^2+\frac{p_r^2}{r^2}+\frac{p_\theta^2}{r^2}\right]\right\}rdrd\theta.\label{eq:freeenergypar}
\end{align}
From this expression, we obtain the components of the molecular field as 
\begin{align}\label{eq:molecularfieldradialtangential}
h_{r}&=-\frac{\delta {\cal F}}{\delta p_r}\nonumber\\
&=-\chi p_r+{\cal K}\left[\frac{1}{r}\partial_r(r\partial_r p_r)-\frac{p_r}{r^2}\right]\\
h_{\theta}&=-\frac{\delta {\cal F}}{\delta p_\theta}\nonumber\\
&=-\chi p_\theta+{\cal K}\left[\frac{1}{r}\partial_r(r\partial_r p_\theta)-\frac{p_\theta}{r^2}\right].
\end{align}

After expressing the radial and azimuthal components of the polarization field as $p_r=S\cos(\psi)$ and $p_\theta=S\sin(\psi)$, the components of the molecular field parallel and perpendicular to the polarization field, $h_\parallel=\cos(\psi)h_r+\sin(\psi)h_\theta$ and $h_\perp=-\sin(\psi)h_r+\cos(\psi)h_\theta$,  read
\begin{align}
h_{\parallel}&=-\chi S+{\cal K}\left[\partial_{rr} S+\frac{\partial_r S}{r} -\frac{S}{r^2}-S(\partial_r \psi)^2\right]\label{eq:molecularfieldpara}\\
h_{\perp}&={\cal K}\left[S \partial_{rr}\psi +\frac{S\partial_r \psi}{r}+2(\partial_{r} S)(\partial_{r} \psi) \right].\label{eq:molecularfieldperp}
\end{align}

 \section{Numerical integration scheme}
 \label{sec:numerics}
 
 The numerical solutions for the steady-state polarization $\mathbf{p}$ and the azimuthal velocity component $v_\theta$ presented in Sects.~\ref{sec:spiral} and \ref{sec:monolayers} were obtained by solving the time-dependent form of Eqs.~\eqref{eq:dinamicadirectorparticular1} and  \eqref{eq:dinamicadirectorparticular2} for $\mathbf{p}$, that is,
\begin{align}
\partial_t S&=\frac{h_\parallel}{\gamma}-\nu S v_{r\theta}\sin(2\psi)\label{eq:dinamicadirectorparticularnum1}\\
\partial_t \psi&=\frac{h_\perp}{\gamma}+S v_{r\theta}\left(1-\nu \cos(2\psi)\right) ,\label{eq:dinamicadirectorparticularnum2}
\end{align}
as well as the time-independent Eq.~\eqref{eq:forcebalance2} for $v_\theta$ with boundary conditions Eqs.~\eqref{eq:boundarycon1}-\eqref{eq:boundarycon3} and \eqref{eq:boundarycon5}-\eqref{eq:boundarycon7}. 
 
These equations were discretized in space with a number of lattice sites of $10^5$. Spatial derivatives were approximated by central finite differences. At a time $t$, the profiles for the polar order $S$ and the angle $\psi$ were first used to compute $v_\theta$ at this time by directly inverting the linear operator. We then used a semi-implicit Euler method to compute $S(r,t+\Delta t)$ and $\psi(r,t+\Delta t)$. Here, the time step $\Delta t$ was chosen such that the maximal relative changes in $S$ and $\psi$ were smaller than $0.01\%$. This procedure was iterated until steady state was reached. We used a random initial condition.

\end{document}